\documentstyle{mn}

\newif\ifAMStwofonts

\ifoldfss
  \ifCUPmtlplainloaded \else
    \NewTextAlphabet{textbfit} {cmbxti10} {}
    \NewTextAlphabet{textbfss} {cmssbx10} {}
    \NewMathAlphabet{mathbfit} {cmbxti10} {} 
    \NewMathAlphabet{mathbfss} {cmssbx10} {} 
  \fi
  \ifAMStwofonts
    \ifCUPmtlplainloaded \else
      \NewSymbolFont{upmath} {eurm10}
      \NewSymbolFont{AMSa} {msam10}
      \NewMathSymbol{\upi}     {0}{upmath}{19}
      \NewMathSymbol{\umu}     {0}{upmath}{16}
      \NewMathSymbol{\upartial}{0}{upmath}{40}
      \NewMathSymbol{\leqslant}{3}{AMSa}{36}
      \NewMathSymbol{\geqslant}{3}{AMSa}{3E}

      \let\leq=\leqslant 
      \let\geq=\geqslant 
    \fi
  \fi
\fi 

\ifnfssone
  \newmathalphabet{\mathit}
  \addtoversion{normal}{\mathit}{cmr}{m}{it}
  \addtoversion{bold}{\mathit}{cmr}{bx}{it}
  \newmathalphabet{\mathbfit} 
  \addtoversion{normal}{\mathbfit}{cmr}{bx}{it}
  \addtoversion{bold}{\mathbfit}{cmr}{bx}{it}
  \newmathalphabet{\mathbfss} 
  \addtoversion{normal}{\mathbfss}{cmss}{bx}{n}
  \addtoversion{bold}{\mathbfss}{cmss}{bx}{n}
  \ifAMStwofonts
    \ifCUPmtlplainloaded \else
      \UseAMStwoboldmath
      \makeatletter
      \new@mathgroup\upmath@group
      \define@mathgroup\mv@normal\upmath@group{eur}{m}{n}
      \define@mathgroup\mv@bold\upmath@group{eur}{b}{n}
      \edef\UPM{\hexnumber\upmath@group}
      \new@mathgroup\amsa@group
      \define@mathgroup\mv@normal\amsa@group{msa}{m}{n}
      \define@mathgroup\mv@bold\amsa@group{msa}{m}{n}
      \edef\AMSa{\hexnumber\amsa@group}
      \makeatother
      \mathchardef\upi="0\UPM19
      \mathchardef\umu="0\UPM16
      \mathchardef\upartial="0\UPM40
      \mathchardef\leqslant="3\AMSa36
      \mathchardef\geqslant="3\AMSa3E

      \let\leq=\leqslant 
      \let\geq=\geqslant 
    \fi
  \fi
\fi 

\ifnfsstwo
  \DeclareMathAlphabet{\mathbfit}{OT1}{cmr}{bx}{it}
  \SetMathAlphabet\mathbfit{bold}{OT1}{cmr}{bx}{it}
  \DeclareMathAlphabet{\mathbfss}{OT1}{cmss}{bx}{n}
  \SetMathAlphabet\mathbfss{bold}{OT1}{cmss}{bx}{n}
  \ifAMStwofonts
    \ifCUPmtlplainloaded \else
      \DeclareSymbolFont{UPM}{U}{eur}{m}{n}
      \SetSymbolFont{UPM}{bold}{U}{eur}{b}{n}
      \DeclareSymbolFont{AMSa}{U}{msa}{m}{n}
      \DeclareMathSymbol{\upi}{0}{UPM}{"19}
      \DeclareMathSymbol{\umu}{0}{UPM}{"16}
      \DeclareMathSymbol{\upartial}{0}{UPM}{"40}
      \DeclareMathSymbol{\leqslant}{3}{AMSa}{"36}
      \DeclareMathSymbol{\geqslant}{3}{AMSa}{"3E}

      \let\leq=\leqslant 
      \let\geq=\geqslant 
    \fi
  \fi
\fi 

\ifCUPmtlplainloaded \else
  \ifAMStwofonts \else 
    \def\upi{\pi}
    \def\umu{\mu}
    \def\upartial{\partial}
  \fi
\fi

\title{PKS2250-41 and the r\^ole of jet-cloud interactions in powerful
radio galaxies.} 

\author[Villar-Mart\'\i n {\it et al.}]
       {M. Villar-Mart\'\i n,$^{1,2}$ 
  C. Tadhunter,$^2$ R. Morganti,$^3$ D. Axon $^4$, A. Koekemoer$^5$
 \\
       $^1$Institut d'Astrophysique de Paris, 98 bis Blvd. Arago, F75014
Paris \\ $^2$Department of Physics and Astronomy, University of
Sheffield, Sheffield S3~7RH, UK\\ $^2$Istituto di Radioastronomia, Via Gobetti 101, 40129 Bologna, 
Italy\\ $^{3}$Department of Physics and Astronomy, University of Manchester, Oxford
Road, Manchester, UK\\ $^{4}$Space Telescope Science Institute, 3700 San Martin
Drive, Baltimore, MD21218, USA}
\date{Accepted 1999 February 25.
      Received .
}

\pagerange{\pageref{firstpage}--\pageref{lastpage}}
\pubyear{1999}

\begin{document}

\maketitle

\label{firstpage}

 \begin{abstract} 
 We present high resolution, long-slit spectra of the jet-cloud interaction
in the powerful southern radio galaxy PKS2250-41. 
We have resolved the emission lines into two main kinematic
components: a broad component (FWHM$\geq$900 km s$^{-1}$) and
a na\-rrow component (FWHM$\leq$150 km s$^{-1}$).  
While the broad component is characterized by a  low ionization level (with particularly
weak HeII$\lambda$4686 emission) and is spatially associated with the radio
lobe, the narrow component is characterized by a higher ioni\-zation level and extends 
well beyond
the radio lobe. Crucially, we 
measure a higher electron temperature for the broad component
($T\sim$30,000 K) than for the narrow
component ($T\sim$15,000 K). The general line ratios and
physical conditions of the two components are consistent
with a model in which the broad component represents gas cooling behind
the shock front driven by the radio jets, while the narrow
component represents the AGN- or shock-photoionized precursor gas. However,
uncertainties remain about the gas acceleration
mechanism behind the shock front: unless the radio components are expanding 
unusually fast in this source, it is likely that entrainment of the warm clouds
in the hot post-shock wind
or radio plasma is required in addition to the initial acceleration across
the shock front, in order to explain the large line widths of the broad component. 
 
	The similarities between the kinematic properties of PKS2250-41 and
some high redshift radio galaxies suggest that the
ambient and the shocked gas have also been resolved in the more
distant objects. 
Given the evidence that the
emission line processes are affected by the interactions between the radio
and the optical structures,
care must be taken when interpreting the UV spectra of high redshift 
radio galaxies.

 \end{abstract}

\begin{keywords}
galaxies: active - galaxies: jets - galaxies: individual (PKS2250-41)
\end{keywords}

\section{Introduction}

	The study of Seyfert galaxies and low redshift radio galaxies
has revealed extended emission line properties which are well explained 
in terms of an illumination model ({\it e.g.} Fosbury 1989)
in which  the ambient gas is illuminated and photoio\-nized
by 
active galactic nuclei (AGN)  in the cores of the galaxies. This
 scenario  has
the   advantage of being in agreement with the unification models 
({\it e.g.} Barthel 1989) in which it is proposed that     many active galaxies  are 
intrinsically identical but appear different
due to orientation effects. 

	However, AGN illumination is not the whole story.  HST images
of Seyfert galaxies (Capetti {\it et al.} 1996) have shown  a close
association  between the narrow-line region emission line structures and
the radio structures. The emission line profiles also reveal disturbed
kinematics and high velo\-city motions ({\it e.g.} Axon {\it et al.} 1998). These 
properties show that strong
interactions are taking place between the radio-emitting plasma and the emission-line 
gas.
 
	Some nearby radio galaxies and the majority  of  high redshift radio galaxies 
(HzRG) ($z>$0.7)   present   clear evidence for similar interactions. Many HzRG show
highly colli\-mated  UV 
continuum and emission line structures   which are closely aligned
with the radio axis (Chambers {\it et al.} 1987; McCarthy {\it et al.} 1987;
Best, Longair \& Rottgering 1996).  The emission line 
spectra show a highly disturbed kine\-matics which  cannot be explained in
terms of purely gravitational motions  ({\it e.g.} Mc.Carthy {\it et al.} 1996,
R\"ottgering {\it et al.} 1997) but suggest that the gas is interacting with
the radio-emitting components. Recently, work by the Stromlo
group has provided a theoretical framework for understanding the emission
line spectra produced by warm emission line clouds involved in jet-cloud
interactions\footnote{Throught this paper we use ``jet-cloud
interaction'' as a generic term to denote interactions between the
radio-emitting components --- including the radio jets, hotspots and lobes ---
and the warm emission line clouds.} ({\it e.g.} Sutherland, Bicknell \& Dopita
1993; Dopita \& Sutherland 1996).

Jet-cloud interactions are therefore a common phenomenon in active
galaxies and can have a strong influence on the properties we observe. The way 
the interactions take place is, however, poorly understood and many questions
remain open. In particular, the gas acceleration mechanism in jet-cloud
interactions remains uncertain, and the extent to which the jet-induced
shocks ionize the emission line clouds is not yet clear.  
 
We are carrying out a project whose main goal is the
study of the  
interaction between the radio and optical structures in radio galaxies. 
The very large distances of HzRG make this study difficult due to the small angular 
sizes, the faintness of the objects, and the fact that optical
observations sample the UV rest frame, where the emission line
physics are highly uncertain.
Therefore we are concentrating
on radio galaxies at intermediate redshifts which are much easier to study.

		PKS2250-41 ($z$=0.308) is one of the few relatively nearby
 radio galaxies with clear evidence for jet-cloud interactions. Due to its proximity 
and the strength of its  interaction,  
PKS2250-41  is  an excellent target  to study   the physics of this phenomenon in 
powerful radio galaxies. 
What we learn from PKS2250-41 will 
provide important clues about what is happening in more distant radio
galaxies. The results will   be useful to understand the r\^ole played
by  interactions between the radio- and optical-emitting structures in active galaxies 
in general. 

\subsection{The  jet-cloud interaction in PKS2250-41. Summary of
previous results.}

   Clark {\it et al.} (1997) and Tadhunter
{\it et al.} (1994) showed that the interaction between the radio jet and the emission
line gas in PKS2250-41 has the following consequences:

\begin{itemize}

\item  Morphological associations between the radio and optical structures: The dominant 
feature of the emission line images ([OIII]$\lambda$5007) is an arc-shape structure 
located at $\sim$5.5 arcsec
($\sim$33 kpc)\footnote{$H_0 = 50$ km s$^{-1}$ Mpc$^{-1}$, $q_0 = 0.0$ assumed 
throughout,
giving an angular scale of 6.0 kpc arcsec$^{-1}$}
west of the nuclear continuum centroid. This arc circumscribes 
the Western radio lobe.

\item  Complex kinematics with    
relatively broad low ionization lines. An anticorrelation is observed
between  line width and the ionization state in the Western arc.
This effect may be
explained, as in SN remnants (Greidanus \& Strom 1992), as    due
to the fact that low ionization  emission lines
are produced mainly in the shocked, compressed  and accelerated gas, while the high 
emission
lines are mainly emitted by the precursor gas.

\item High pressures   measured in the western
radio lobe, were the interaction is strongest. This suggests
that the radio plasma and the line emitting gas are in approximate pressure
equilibrium. 

\end{itemize}
   
The authors propose a scenario in which a direct collision
between the radio jet and a companion galaxy in the surrounding group
is taking place. 
 
In this paper we present new long-slit spectra for PKS2250-41,  
which cover a larger spatial
region than stu\-died in the previous work,  and contain information not 
only for the regions along the radio axis, but also
perpendicular to it. The high spectral
resolution of these data has allowed us to study in detail the    
properties and origin(s) of the diffe\-rent kinematic components revealed in
the strong emission lines. 

\section[]{Observations}

 \begin{table*}
\normalsize
\centering
\caption{Log of the spectroscopic observing run.}
\begin{tabular}{ccccccc}
\hline
 
Central $\lambda$ &  Grating &  $t_{exp}$  &   $\Delta\lambda$ & Resol.(\AA)  & 
Seeing(")	& Slit(") \\	\hline
	&	&  	& Radio axis (PA270)	&	&	&	\\
4883 	&  1200V  & 3600    & 4485-5300  & 1.7$\pm$0.1  & 0.6     &  1.5     \\ 	
6250 	&   600R & 5400    & 5415-7040  &   3.0$\pm$0.3  & 1.0    &  1.5    \\ 	
6560	&  1200V   &  5400 & 6165-6980    &  1.4$\pm$0.1 & 0.6   &  1.5  \\ 
8300 	&  600R   & 6000     & 7520-9130 & 3.3$\pm$0.3 & 1.0     & 1.5     \\  \hline	
	&	&	& 2.5 arc sec S	&	&	&	\\
6560	&  1200V   &  5400 & 6165-6980    &  1.4$\pm$0.1 & 0.6   &  1.5  \\ \hline
	&	&	& 2 arc sec N	&	&	&	\\
6560	&  1200V   &  5400 & 6165-6980    &  1.4$\pm$0.1 & 0.6   &  1.5  \\ \hline
	&	&	& Cross cut (PA0)	&	&	&	\\
6560	&  1200V   &  5400 & 6165-6980    &  1.4$\pm$0.1 & 0.6   &  1.5  \\
\hline	
\end{tabular}
\end{table*}
The spectroscopic observations were carried out on the nights 1994 September 27/28 
using   the Royal 
Greenwich Observatory (RGO) spectrograph 
on the Anglo Australian Telescope. The detector was a  Tek CCD with 226$\times$1024 
pi\-xels of 27 $\mu$m$^2$, resulting in a
spatial scale of 0.81 arcseconds per pixel.

  The data were taken with the slit
approximately parallel (PA270) and perpendicular (PA0) to the radio
axis  using both intermediate  (3.0 -- 3.3\AA ~FWHM) and
high (1.4 -- 1.7\AA ~FWHM) spectral resolutions. We covered most of the 
projected area of the Western arc
by obtaining spectra along three slit positions at PA270: along the radio axis, 2.5 arc 
sec South
(2.5"S slit hereafter) 
and 2 arc sec North (2"N slit hereafter). The slit positions are presented in Fig.~1. 
A log of the spectroscopic observations is shown in Table 1.

 \begin{figure}
\includegraphics{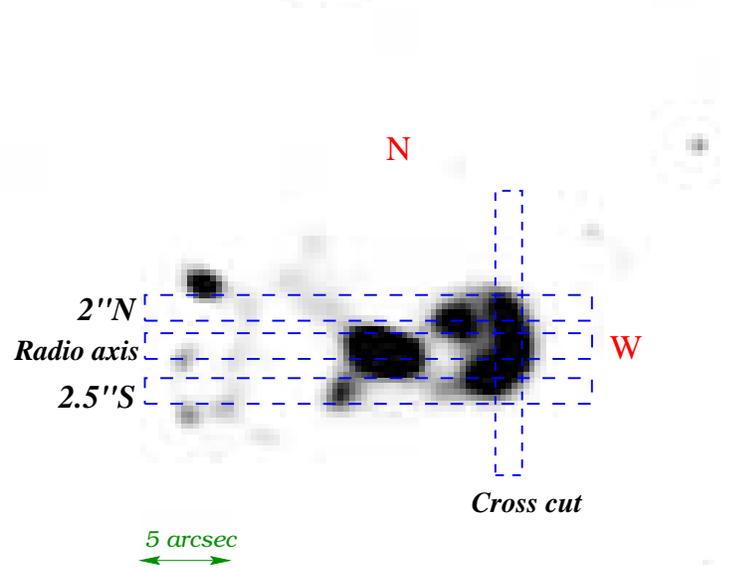}
\vspace{3.5in}
\caption{[OIII] narrow band image of PKS2250-41 (from Clark {\it et al.} 1997)
with the slit positions overplotted.}
\end{figure} 

We did the same with the overlaps between the cross cut and  the 2.5"S and 2"N slits. We 
detect (as expected) three kinematic components (two narrow+one broad) at the radio axis 
position. 
We have isolated the redder narrow
component and plotted its properties
in Fig.~5 (solid triangles).  The good agreement with the cross cut values
for the narrow component  confirms that {\it the blue component is not detected
N and S of the radio
axis along PA0}.

  The data reduction was carried out using standard methods provided in IRAF
(a detailed description can be found in  Villar-Mart\'\i n et al. 1998).

 \section{Analysis and results}

\subsection{The fitting procedure}

 Both IRAF and STARLINK (DIPSO) routines were used to measure the
emission line fluxes, line widths and line centers.  
For the blends, decomposition procedures were used in
STARLINK (DIPSO), fitting several Gaussians at the expected positions of the
components.  As the number of possible mathe\-matical solutions is large, we applied
theoretical constrains when necessary (like fixed ratios between line fluxes
or separation in wavelength).  In some cases, the information obtained from
the strong [OIII]$\lambda\lambda$5007,4959 lines was used to constrain the input
parameters for the fits to other lines. We assumed, for instance,  that
all lines have the same number of kinematic
components as those discovered in the [OIII] lines. Sometimes
we also constrained the kinematic components to have the same velocity
width (taking into account instrumental broadening and wavelength
dependence of velocity) as the [OIII] components.
\begin{figure*}
\includegraphics{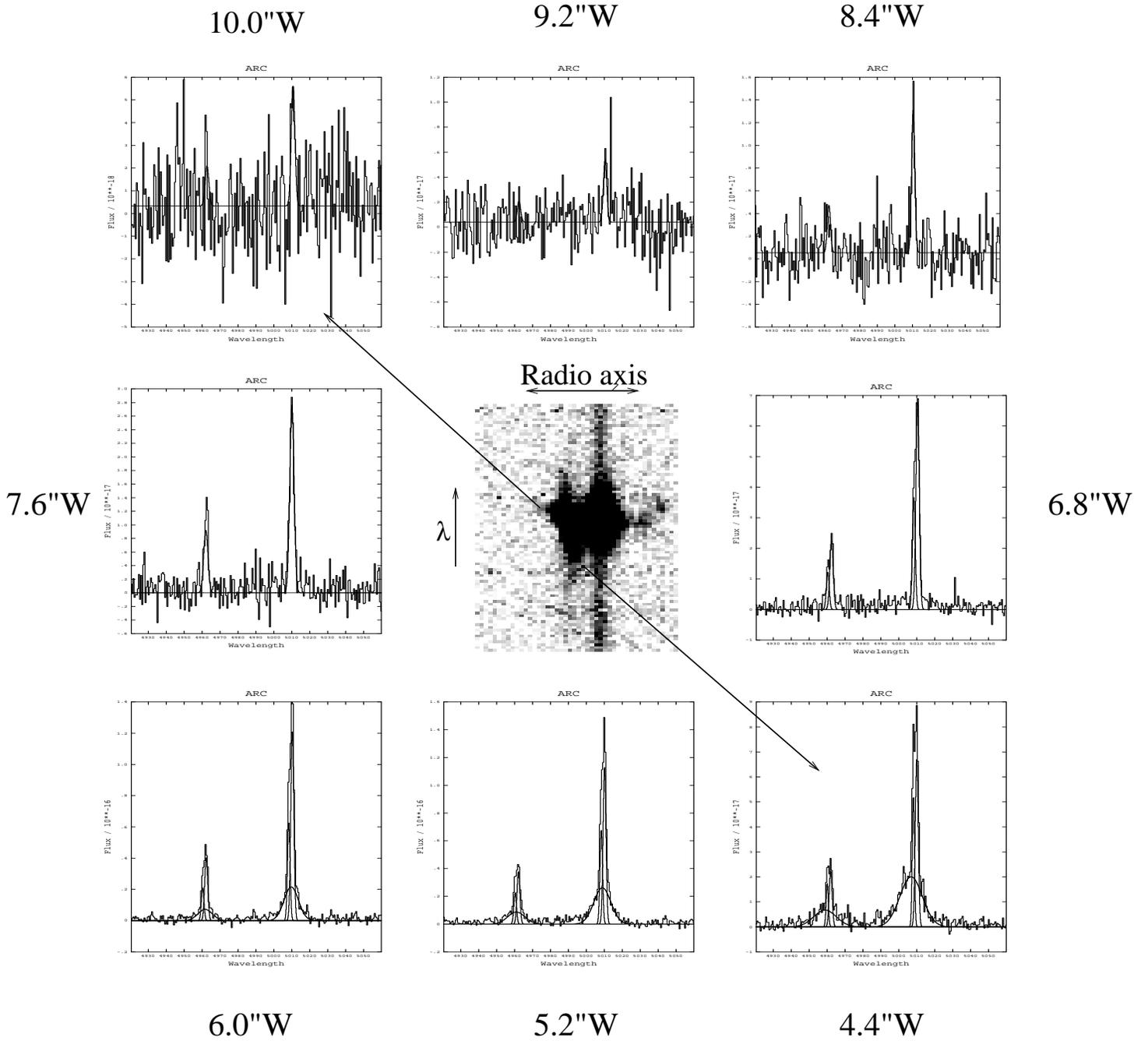}
\vspace{8in}
\caption{The fitting procedure: Example of the spectral decomposition procedure done for 
PKS2250-41. The central panel shows the 2-D [OIII]$\lambda$5007 spectra along the radio 
axis. Dipersion in $\lambda$ varies in the vertical direction and space in horizontal
direction. The plots  show the 1-D spectra
 created from every column.  The emission lines of every spectrum
were fitted with the necessary kinematic components.  The smooth solid
 lines indicate the 
results of the fits. The inner and
outer   pixels
in the arc have been indicated with arrows. The position of each spectrum
relative to the nucleus of the galaxy is also indicated.}
\end{figure*}
	
	The [OIII]$\lambda\lambda$5007,4959
and H$\beta$ lines were used to study the spatial variation of the properties
of the different kinematic components revealed by the fitting procedures.
In order to do this,  we   extracted a 1-D spectrum for
each spatial pixel across the Western arc the region
of greatest interest to us,  which shows the 
clearest  signs of jet-cloud interactions. 
Those pixels for which there was a risk of contamination 
by the  emission from the near-nuclear regions of the host galaxy
were
excluded.    As an example of the fitting procedure, we present in Fig.~2 the results of 
the fits to the [OIII] lines for spatial pixels
across the Western arc  (slit aligned along PA270).

Apart from the [OIII]$\lambda\lambda$5007,4959
and H$\beta$ lines, most of the lines are  faint, and the spatial 
information is difficult to obtain on a pixel-by-pixel
basis for the individual components.
In order to increase the signal to noise ratio and to obtain information on the 
individual kinematic
components, we  spatially integrated 
the emission from the brightest pixels across the Western arc 
(using 1.5$\times$3.2 arcsecond aperture, centered 5.6 arcseconds to the west
of the nucleus along PA270) and
fitted the fainter lines in the integrated
spectrum. The strongest lines in 
each spectrum were used to constrain the input parameters for the fits to
the fainter lines in the same spectrum.

	The result of this procedure is the isolation of the diffe\-rent
kinematic components in each line, characterized by flux, line width
and central wavelength. It is, therefore, possi\-ble to study the kinematics and 
line ratios of each kinematic component separately.

\subsection{Results from the PA270 slits}
  
	~~~~ In this section we analyse the information provided by the 
[OIII]$\lambda\lambda$5007,4959 and H$\beta$ lines in the high 
resolution spectra obtained along PA270 (approximately parallel
to the radio axis).    Our goal is to
isolate the different gaseous components contributing to the line emission
and  study the spatial variation of the kinematic and ionization properties.
The results for the perpendicular slit position (PA0) are presented 
separately in section \S3.3.

 	The fits to the lines reveal  complex kinematics with
the presence of at least three   
spatially extended components: two narrow components (FWHM$\sim$60-200 km s$^{-1}$) 
separated by
$\sim$140 km s$^{-1}$ and a broad component
($\sim$500-900 km s$^{-1}$). We will distinguish the two narrow components by
labeling them as  
the blue narrow component (component at shorter wavelength) and the red narrow component 
(component at longer wavelengths).

  We plot  the spatial variation of  the kinematic, flux   and ionization properties of 
the
narrow components in Fig.~3. The properties of the broad components are
presented in Fig.~4. In order to highlight the possible connection with the radio 
structures, we have indicated in these diagrams the position of the radio hot 
spot and the outer
edge of the radio lobe.

\vspace{0.15cm}

 \centerline{\it The narrow components (Fig.~3)}

\vspace{0.15cm}

	Fig.~3 shows that all the narrow components
have simi\-lar line widths across the  full extent of the nebulosity
(FWHM $\sim$60-200 km s$^{-1}$).    
The velocity shift with respect to the
nuclear emission is rather constant for each component.  The blue narrow component --- 
detected along the radio axis and also in the detached
emission line region $\sim$4.5 arcse\-conds to the NW of the nucleus
(see Figure 3)  ---
is noticeably displaced in velocity with respect to the other components
(velocity shift: 80-140 km s$^{-1}$).

\begin{figure*}     
\includegraphics{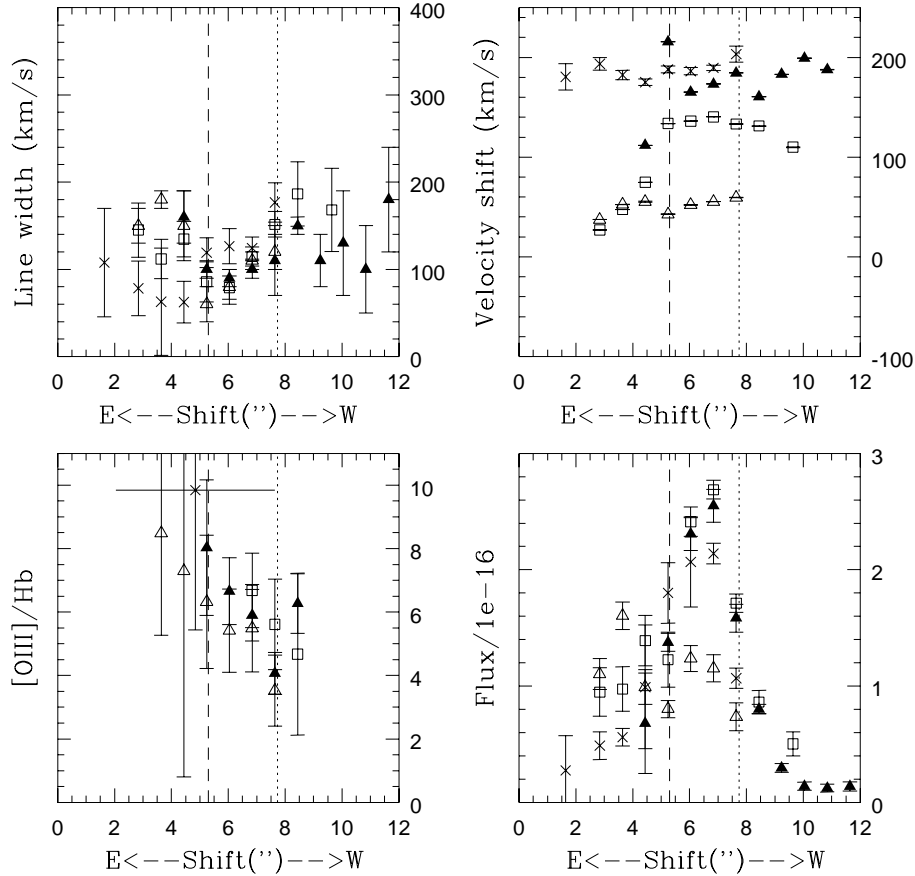}
\vspace{5in}
\caption{{\it The narrow line emission along the slits parallel to the
radio axis:} Spatial variation of the kinematic, flux and   ionization properties of the 
narrow
components detected across the Western arc  along
the  slit positions parallel to the radio axis. The spatial shift has been
calculated with respect to the continuum centroid of the galaxy.
{\it Panels: }Top left:  FWHM. Top right: Velocity shift with respect to the nuclear 
emission. Bottom left: 
Ionization level. Bottom right: Fluxes. 
\hspace{1.cm}
{\it Symbols}: Radio axis  red component -$>$   solid  triangles. Radio axis 
blue component -$>$   open  triangles. 2.5"S slit  -$>$ x symbols.
2"N slit -$>$ open squares. The   dashed and the dotted vertical lines mark the 
positions of the hot spot and the edge of the radio lobe respectively 
at the slit located along the radio axis. Note the similarities
of the narrow line emission along the 2.5"S and 2"N slits, and the redder narrow
component along the radio axis.}
\end{figure*}

	As revealed by the
[OIII]/H$\beta$ ratio, the ionization level of the narrow components  
shows little variation, both in  absolute value and in spatial 
variation, across the arc.  Due to the faintness of H$\beta$ it was necessary
to add all the   pixels along the 2.5"S slit.  The integrated value 
for this slit 
fits in with  the variation defined by the other narrow 
components. 
   
	 The flux curves (Fig.~3 bottom right) of the redder
na\-rrow  component along the radio axis and the narrow component  along
the extreme slit positions are almost identical. The
absolute flux values also remain rather constant when we move 
North and South from the radio axis. The   redder narrow 
component extends far beyond the hot spot
(6.5 arc sec)  and $\sim$4.8 arc sec
beyond the edge of the radio lobe. 

In contrast, the flux curve of the narrow blueshifted 
component detected along the
radio axis is different in the sense that
{\it this component disappears abruptly at the edge of the radio
lobe.} 

\vspace{0.15cm}

\centerline{\it The broad components. (Fig.~4)}

\vspace{0.15cm}

The broad component presents a strong variation in its kinematic,
ionization and   flux properties along the radio axis. The flux {\it peaks at the 
position of  the hot spot},
where its emission dominates over the narrow components. {\it It is also at this
position that the line is broadest} (FWHM$\sim$900 km s$^{-1}$). 

	At other positions along the radio axis and the extreme slits, the
FWHM and flux  
are rather constant. The velocity curves   are very similar along the
radio axis and the 2"N slit, with an S-shape, although spatially shifted ($\sim$1.6 arc 
sec).
   
 \begin{figure*}  
\includegraphics{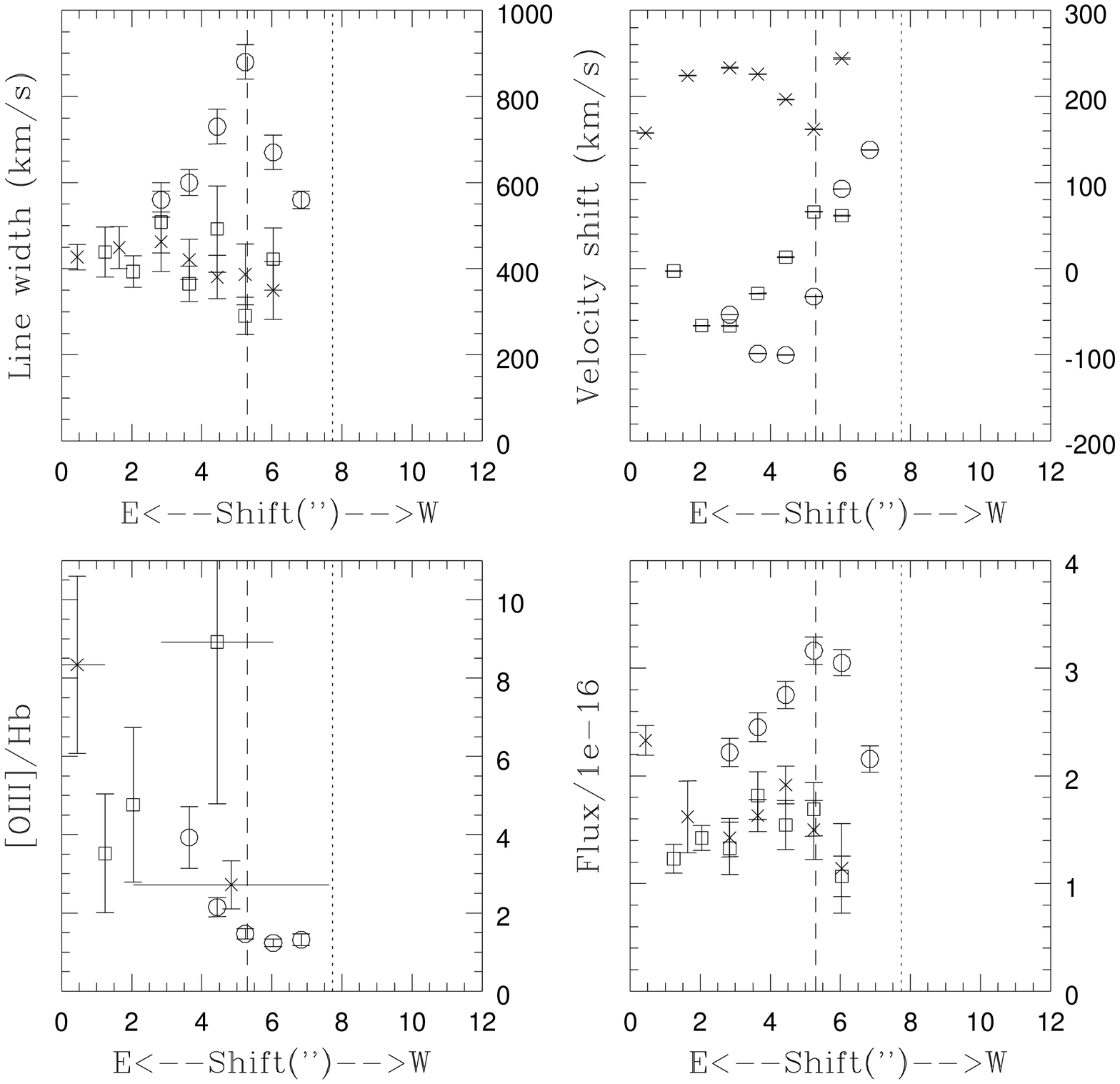}
\vspace{5in}
\caption{{\it The broad line emission along the slits parallel to the
radio axis:} 
Radio axis  -$>$  open circles. 2.5"S slit  -$>$ x symbols.
2"N slit -$>$ open squares. Note the maximum of the
FWHM and the flux of the broad component at 
the position of the hot spot. Note also the low ionization level of this
component.}
\end{figure*} 
	{\it The ionization level of the broad component is  low
compared to the narrow components} ([OIII]/H$\beta\sim$1.5-4). As observed for the 
narrow component,
the ionization level increases towards the galaxy along the radio axis.

 It was necessary to bin together several pixels to increase the S/N ratio along the 
2.5"S and 2"N slits. The error is still
too large for the measurement on the 2"N slit, but the result obtained from
the 2.5"S spectrum  is
consistent with   a very low ionization level of the broad component
([OIII]/H$\beta\sim$2.8), 
similar to the value measured along
the radio axis. 
 
\subsection{Results from the cross-cut slit (PA0).}

   	~~~~In this section we analyse the information provided by the 
[OIII]$\lambda\lambda$5007,4959 and H$\beta$ lines in the 
intermediate reso\-lution spectra obtained with the slit  perpendicular to the radio axis.

 	Two spatially extended components are detected: a broad component 
(FWHM$\sim$500-900  km s$^{-1}$)  and a narrow component    
(FWHM $\sim$200 km s$^{-1}$).  The results for both components are plotted in
Fig.~5.

The low spectral resolution of this spectrum prevents us
from resolving  the two narrow components detected along the
radio axis (see \S3.2).   
Therefore, the information  near and on the radio axis
position is  contaminated by the presence of the blue narrow component 
along the radio axis. In order to eliminate this
effect, we have  extracted   a
1-D spectrum  from the high resolution spectrum obtained along the radio axis   and 
corresponding to the spatial
overlap   with the cross cut slit (see Fig.~1).

 \begin{figure*}  
\includegraphics{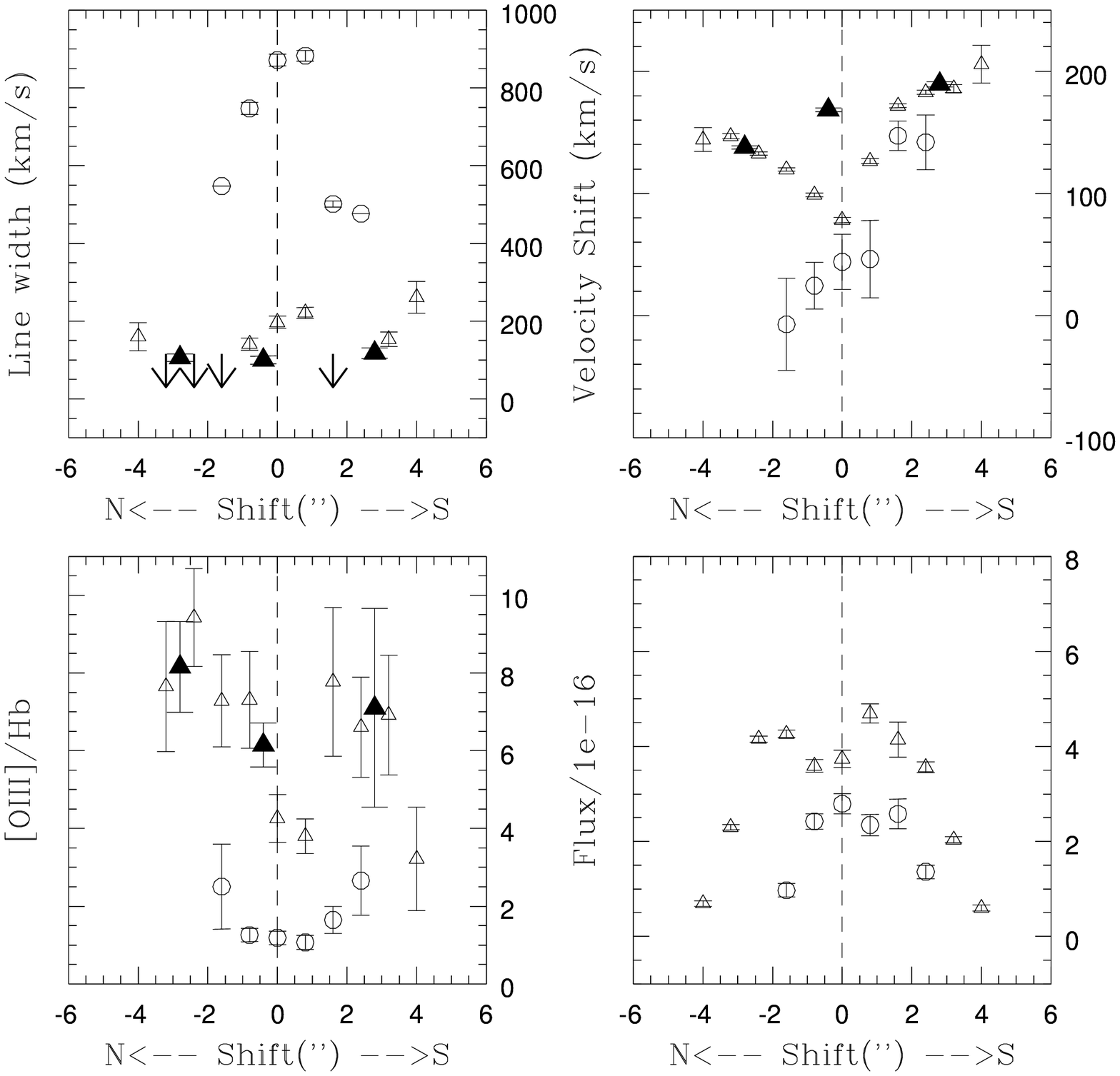}
\vspace{5in}
\caption{Spatial variation of the kinematic, flux and  ionization properties of the arc 
along
the {\it crosscut} slit position. {\it Both} the narrow (open triangles)
and the broad (open circles)
components are plotted. The spatial shift is now calculated with respect
to the radio axis position (dashed vertical line). Note change of symbols:  Broad 
component -$>$ Open circles.  The solid triangles
(see text)
are measurements from the PA270 spectra extracted from the regions
where they overlap with the PA0 spectrum (see Fig.~1).}
\end{figure*}

\vspace{0.15cm}

\centerline{\it The narrow component (Fig.~5)}
 
\vspace{0.15cm}

	The narrow component  is detected across the full extent ($\sim$ 8.9 arc sec) of 
the emission line nebulosity in the N-S direction
along PA0. The FWHM (100-150 km s$^{-1}$), velocity shift (150-200 km s$^{-1}$)
and ionization level ([OIII]/H$\beta$=6.5-8) are {\it rather constant} across
the slit (N-S direction). This is clear once we have corrected for the
contamination of the blue narrow component at the radio axis position.

  	The slight increase on the FWHM
at the position of the radio axis, the V shape of the velocity curve, 
and possibly also the drop in the ionization level  are due to the contamination
by the blue component.  The flux curve  shows also a drop
across two pixels at the radio axis position. The drop must be more pronounced, since 
the
blue narrow component contaminates the emission in these two pixels. This
suggests that the narrow component presents an edge brightening in the direction 
perpendicular to the radio axis. 

\vspace{1cm}

 	\centerline{\it The broad component (Fig.~5)}

\vspace{0.15cm}

	The broad component is spatially extended by 4.9 arcseconds  
N to S across the radio axis.  

  The width varies substantially across the arc
in the N-S direction and {\it peaks at the position of the radio axis}. The width
drops in the outer parts of the arc and is consistent with the values measured along
the 2.5"S and 2"N slits (FWHM$\sim$500  km s$^{-1}$). The flux has a maximum 
near the radio axis position.

 	 The ionization level of the broad
component is remar\-kably low  ([OIII]/H$\beta$=1.2-3) compared to the narrow component, 
as expected from our previous results (see \S3.2)

 \vspace{0.2cm}
   
	In summary, 
disturbed kinematics   is apparent   across the whole 
of the W arc. Multiple components are detected which present different
spatial extensions, kinematic and ionization properties. A summary of the results   is 
presented in Fig.~6.

\subsection{Additional information along the radio axis.}

 	   We have spectra along the 
radio axis covering a wider spectral range, including numerous optical lines which
can provide additional information on the origin and physical pro\-perties (density, 
temperature) of the kinematic components
and their physical properties. 

	After all the 2-D frames were aligned, we  extracted 1-D spectra 
from the brightest spatial pixels of the Western arc
 (1.5$\times$3.2 arcsecond aperture, centered 5.6 arcseconds to the west
of the nucleus along PA270). We then isolated the broad and narrow
components   in all the interesting 
emission lines, using the fitting method  described above.

 	 Table 2  presents the   fluxes of the main optical lines rela\-tive
to H$\beta$ for the two   components revealed by the fits. 
Notice the clear differentiation
between the broad and the na\-rrow components: as the previous results suggest,
strong low ionization lines dominate
the spectrum of the  broad component and high ionization lines dominate
the spectrum of  the narrow component.

\begin{table} 
\normalsize 
\centering
\caption{Line fluxes relative to H$\beta$ of the broad and narrow components
measured along the radio axis in the Western arc. H$\beta$ flux is given
in units of W m$^{-2}$. The measurements were made from
a spectrum extracted using
1.5$\times$3.2 arcsecond aperture, centered 5.6 arcseconds to the west
of the nucleus along PA270.} 
\begin{tabular}{lll} \hline
\hline
	&  Narrow 	& Broad 	\\ \hline
~Flux(H$\beta$)	&    (2.7$\pm$0.1)$\times10^{-19}$    &  (7.1$\pm$0.2)$\times10^{-19}$ 
\\  \hline	
~[OII]$\lambda$3727 &       7.2$\pm$0.3   &  4.7$\pm$0.07 \\ 
 ~H$\gamma$ &      0.45$\pm$0.06   & 0.32$\pm$0.05  \\ 
~[OIII]$\lambda$4363 &   $<$0.1 	&   0.07$\pm$0.01    \\ 
~HeII$\lambda$4686 &      0.22$\pm$0.03   & $<$0.08   \\ 
~[OIII]$\lambda$5007 &      5.4$\pm$0.2    &   1.35$\pm$0.06\\ 
~[NI]$\lambda$5199 &     0.24$\pm$0.04    &  0.22$\pm$0.03  \\
~[OI]$\lambda$6300 &   0.8$\pm$0.1     &  1.10$\pm$0.08 \\ 
~H$\alpha$  &     3.3$\pm$0.2     &  3.1$\pm$0.1 \\ 
~[NII]$\lambda$6583   	&  1.9$\pm$0.1   &  2.63$\pm$0.09  \\ 
~[SII]$\lambda$$\lambda$6716+6732   &   2.2$\pm$0.3   &  3.4$\pm$0.2 \\  
\hline 
\end{tabular}
\end{table}

\begin{figure*} 
\includegraphics{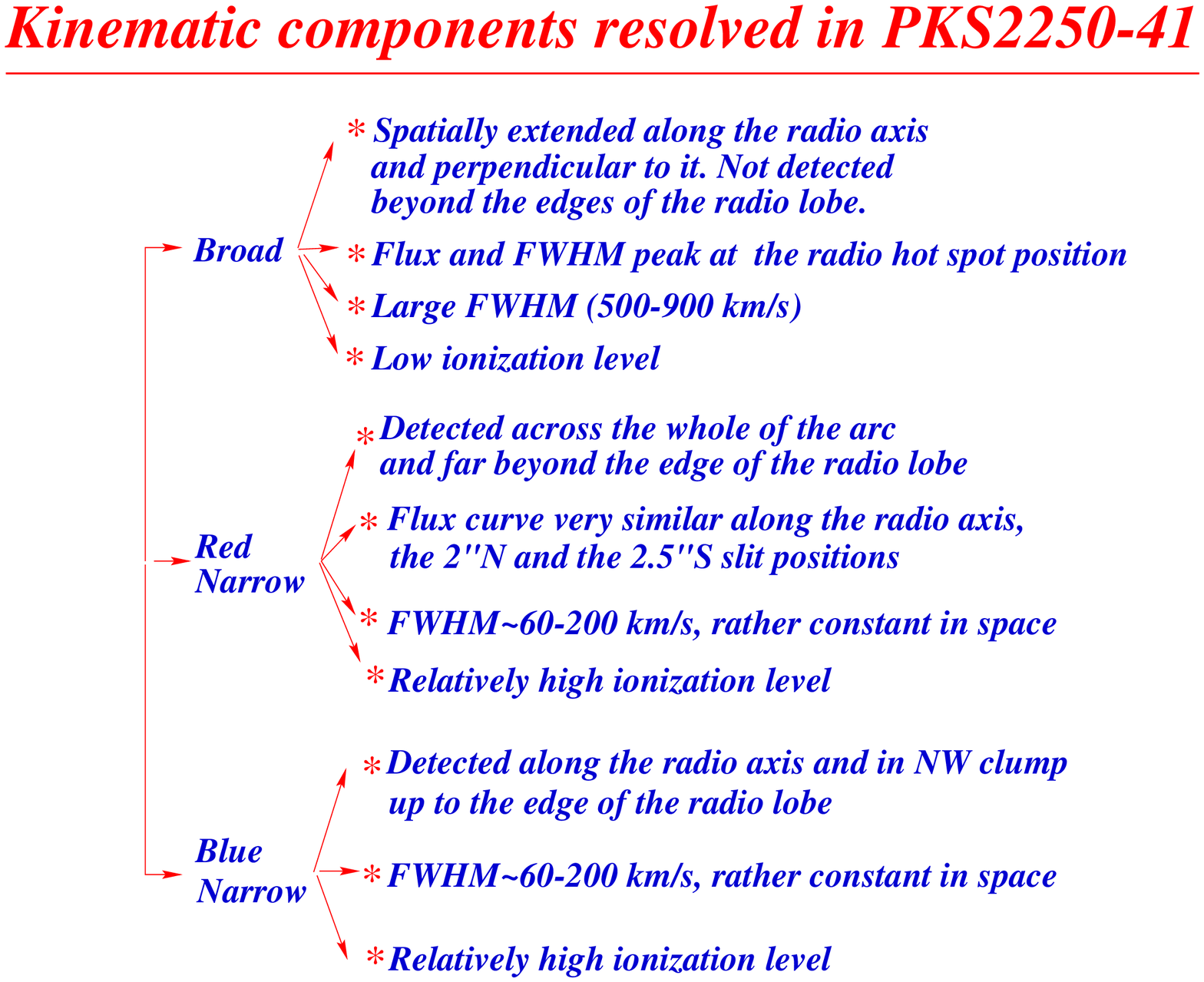}
\vspace{5in}
\caption{Summary of the properties of the three kinematic components detected
in the Western Arc of PKS2250-41.}
\end{figure*}

\begin{table}
\normalsize 
\centering
\caption{Results of the fit to the [OIII]$\lambda$4363 line.
For the radio axis the measurements were made
from a spectrum extracted using a  1.5$\times$3.2 arcsecond aperture
centered 5.6 arcseconds west of the nucleus along PA270, whereas for
the cross-cut a 4.0 $\times$1.5 arcsecond extraction aperture centered 
on the radio axis was used.} 
\begin{tabular}{llll} \hline
\hline
~[OIII]$\lambda$5007/$\lambda$4363	& Narrow	&  Broad \\ \hline
Radio axis &     $>$52      &   17$\pm$5   \\ 
Cross cut  &   50$\pm$25  &   10$\pm$4        \\
Added 	& 50$\pm$20	& 13$\pm$4  	\\
\hline 
\end{tabular}
\end{table}
	Particularly notable are the relative faintness
of HeII$\lambda$4686 and the low values of the 
temperature-sensitive [OIII]$\lambda$5007/[OIII]$\lambda$4363  line ratio 
measured in the
broad component (see below). Although these characteristics
were also noted  by Clark {\it et al.} (1997) on the
basis of single Gaussian fits to low resolution spectra, 
the isolation of the di\-fferent
kinematic components shows that   {\it these are properties
of the broad component}. Thus, it is likely that the ionization minimum
deduced from low resolution spectra at the position of the radio lobe
by Clark {\it et al.} (1997) is a consequence of the increased dominance
of the low ionization broad component at that location. The new results
also suggest that the anticorrelation between
the line width and ionization  stage deduced from single Gaussian fits
to the low resolution spectra is 
a consequence of the low ionization state of the broad component.

\subsubsection{The electron temperature}

	We  have used     the [OIII]$\lambda$4363
and [OIII]$\lambda$5007 lines to measure the electron temperature of the two
main kinematic components.    	
H$\gamma$ and [OIII]$\lambda$4363 were fitted together. 
We have done this analysis for three different spectra:
1)  spectrum of the arc  extracted
along the radio axis from those pixels where the broad component
was brightest (same aperture as in \S3.4) 2)  spectrum extracted along the cross cut 
slit position
also from the pixels where the broad component was brightest (1.5$\times$4.0 arcsecond 
extraction aperture centered 
on the radio axis); and 3) adding
these two spectra after correcting for any possible shift in wavelength.
  	
The best fit was obtained by constraining    the   velocity widths (obtained
from the [OIII]$\lambda\lambda$
5007,4959 lines in the same spectrum) of all the kinematic
components involved.  We show in Fig.~7  the results
of the three fits.

\begin{figure*}
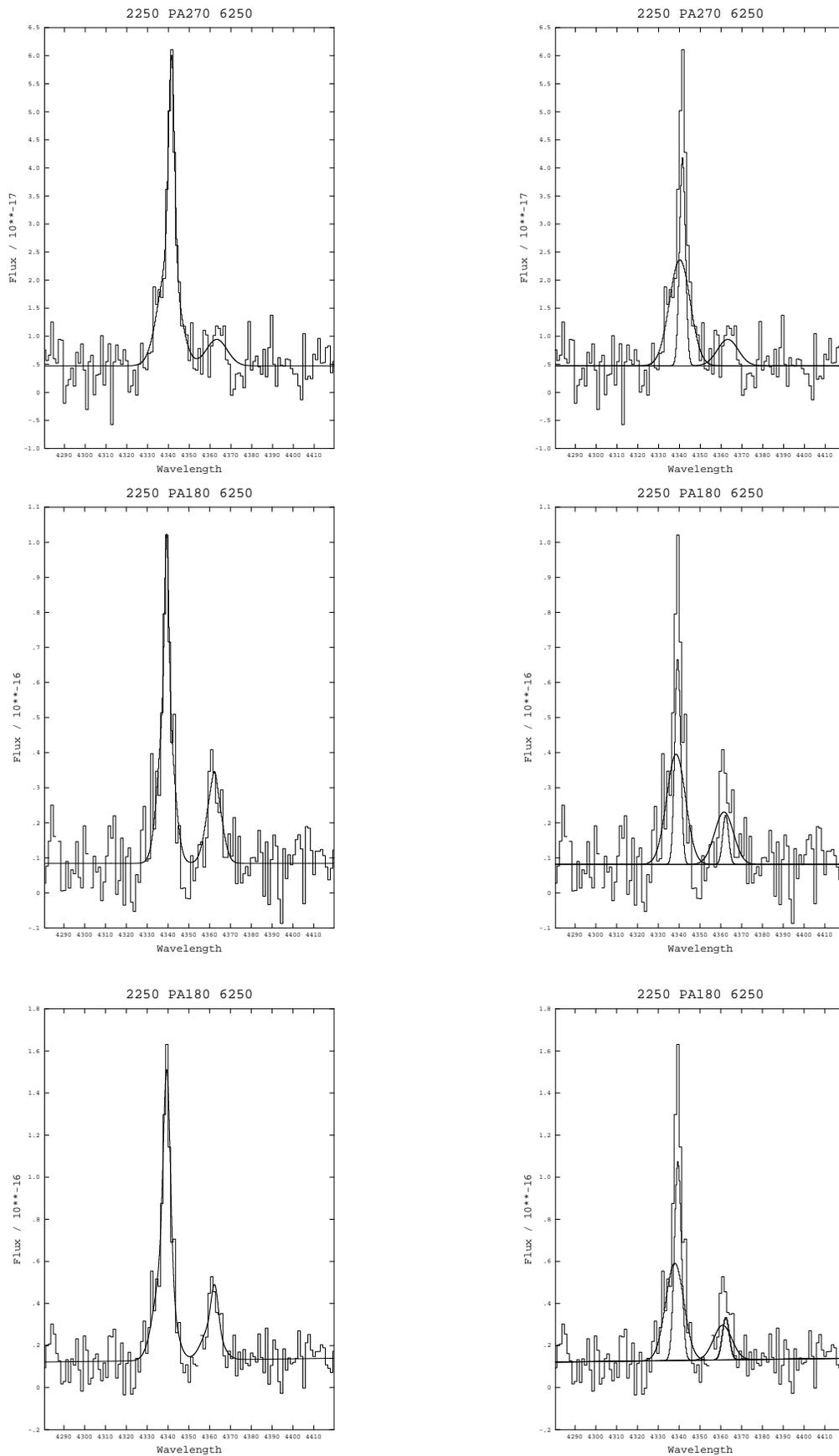
  
\includegraphics{Fig7a.ps}
\includegraphics{Fig7b.ps}
\vspace{3.in}
 \includegraphics{Fig7c.ps}
\includegraphics{Fig7d.ps}
\vspace{3.in}
 \includegraphics{Fig7e.ps}
\includegraphics{Fig7f.ps}
\vspace{3in}
\caption{{\it The fit to the H$\gamma$+[OIII]$\lambda$4363 lines.} 
Top panels:  fit to the spectrum extracted from the radio axis slit.
Middle panels: fit to the spectrum extracted from the cross cut slit position.
Bottom panels: fit to the spectrum sum of the two previous spectra.
Left: Solid smooth line represents the fit to the data.  Right: Solid smooth
lines represent the individual components (narrow and broad).}
\end{figure*}
 
  \begin{figure}
\includegraphics{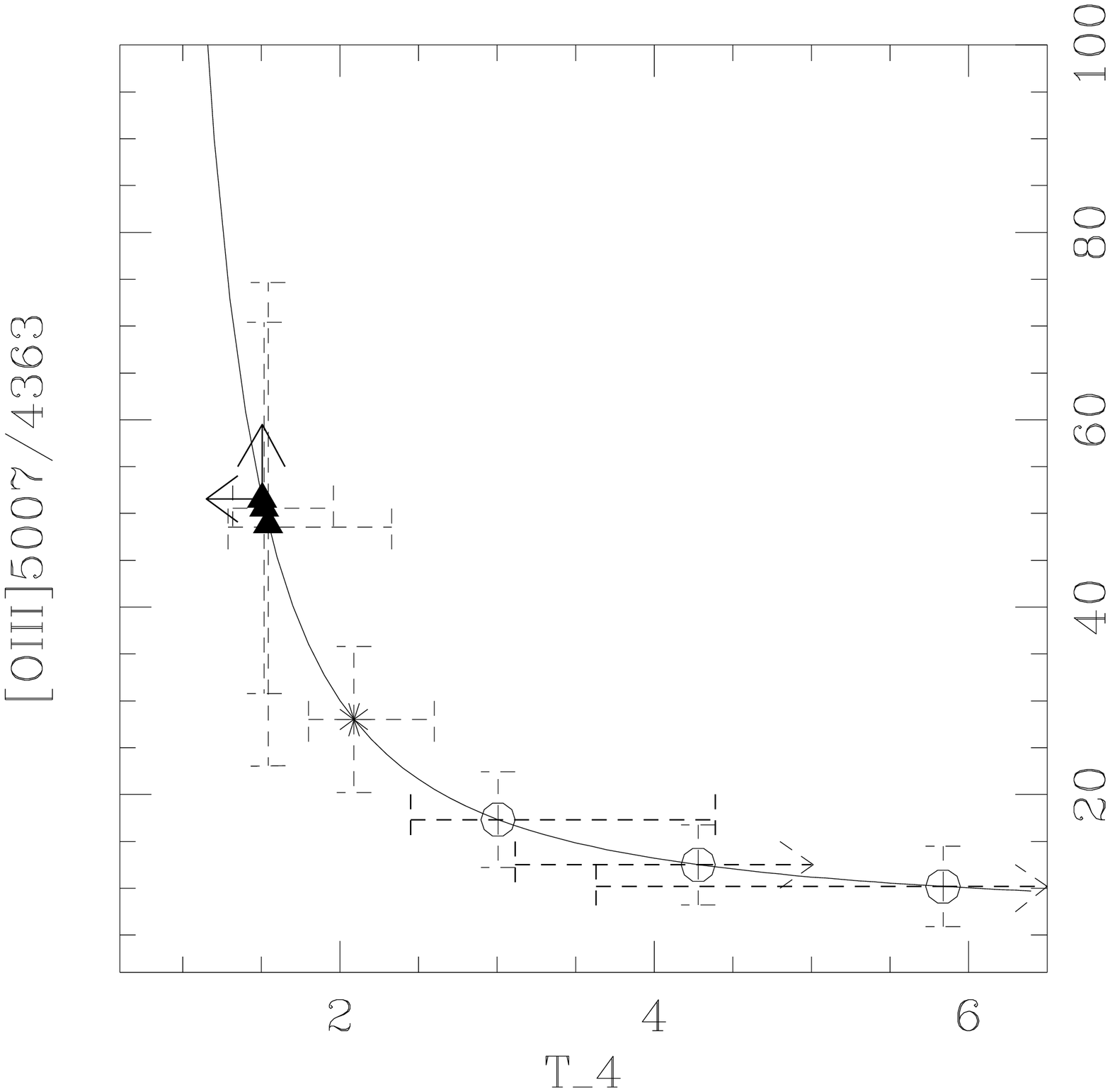}
\vspace{3.5in}
\caption{{\it The temperature}: [OIII]$\lambda$5007/[OIII]$\lambda$4363
{\it vs.} predicted T/10$^4$ temperature.  The results of the fits to
the three spectra (see text) are presented. The broad component is represented 
as an  open circle  and the narrow component  as a solid triangle. The star
represents the measurements from the total flux of the lines. All fits
are consistent with   the broad component being hotter. }
\end{figure} 
Although the [OIII]$\lambda$4363  line is faint, we are confident that the results
are reliable because: 1) the fits to all spectra produce   consistent
results; and  2) the H$\gamma$/H$\beta$ ratios  
are consistent within the errors 
with Case B recombination.

 	We have used the measured    [OIII]$\lambda$5007/$\lambda$4363 ratios
to calculate the electronic temperature of the gas using the formula
in Osterbrock 1989 valid for low densities. 
The densities in the arc (see Clark {\it et al.} and next section) do not exceed a few 
hundred
and the low density formula is valid.
We present  in Fig.~8 the function [OIII]$\lambda$5007/$\lambda$4363=f($T$),
with 
the measured values of the line ratio
for the two kinematic components (for the three fits) plotted. 
The broad component
lies in a region of the diagram where    [OIII]$\lambda$5007/$\lambda$4363
is highly sensitive to the temperature. For this reason the uncertainty
in the temperature is large, in spite of the rela\-tively small errors in the
[OIII]$\lambda$5007/$\lambda$4363 ratio.  It is clear, however, 
that the broad component has a higher electron 
temperature than the narrow component: we
obtain,  $T_e =30,000^{+8,000}_{-4,000}$ K,  $T_e =60,000^{+100,000}_{-25000}$ K, and 
$T_e =41,000^{+10,000}_{-8,000}$ K  
for fits to the radio axis, cross-cut and
co-added spectra respectively, while for the narrow component
we obtain $T_e = 15,000^{+1300}_{-1000}$. Note that, unlike the situation for
the nuclear regions of active galaxies, the equivalent width of
the [OIII]$\lambda$4363 line is large in the regions of interest.
Consequently, systematic effects related to the subtraction of the
underlying stellar continuum are unlikely to be a significant source
of error.

We have also calculated the temperature using the {\it total} flux of the 
lines in the co-added spectrum (with no constraint on the properties of the lines). The 
result  --- [OIII]$\lambda$5007/$\lambda$4363=
28$\pm$8 ---
is consistent with an intermediate temperature between the narrow and 
the broad component (see Fig.~8): $T_e$=21000$^{+5000}_{-3000}$K. This supports the 
consistency of
our results. Assuming that the narrow component is cooler, 
$T_e$=21000 K is a {\em lower
limit} on the temperature of the broad component.

\begin{figure*}[hb]  
\includegraphics{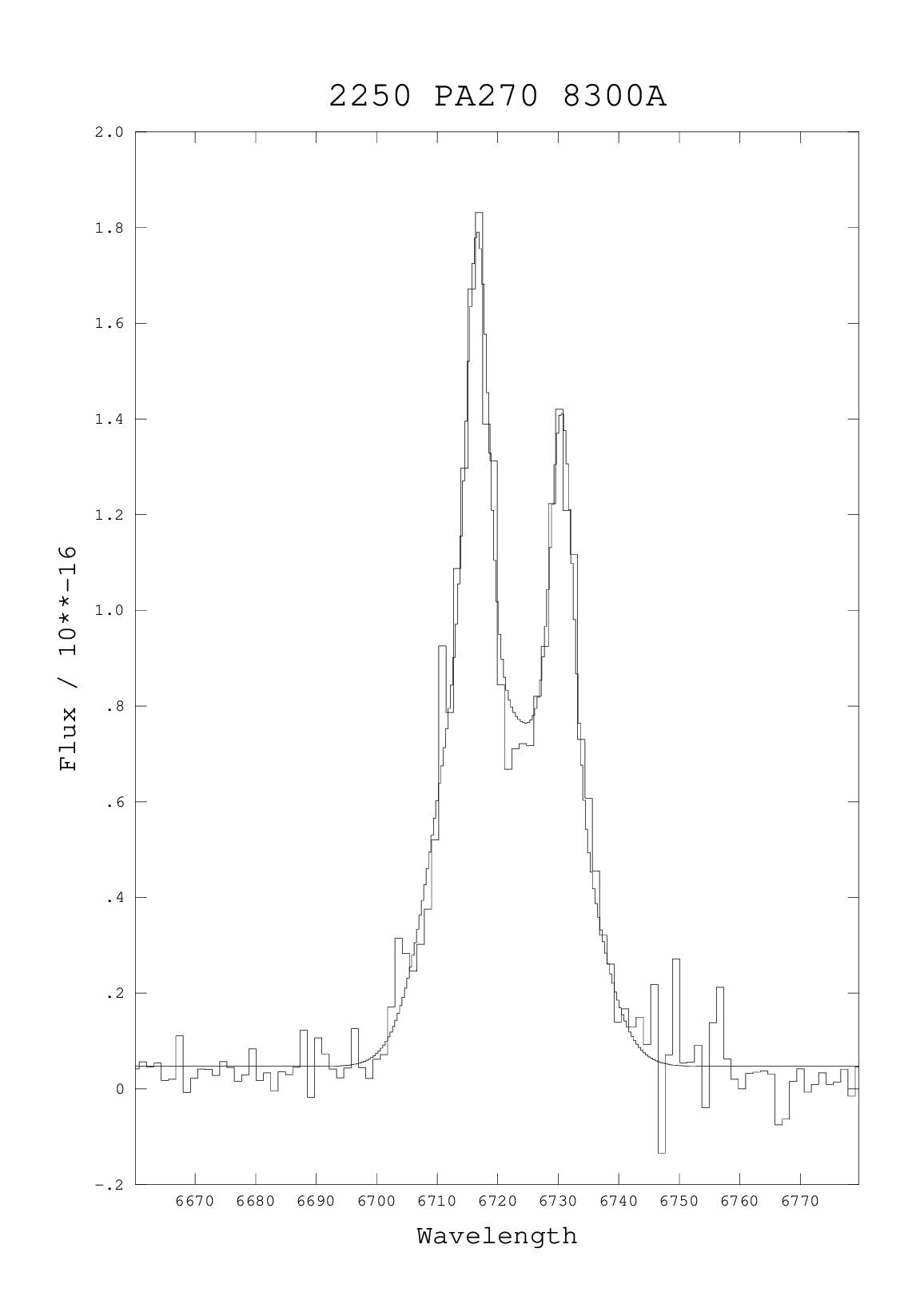}
\includegraphics{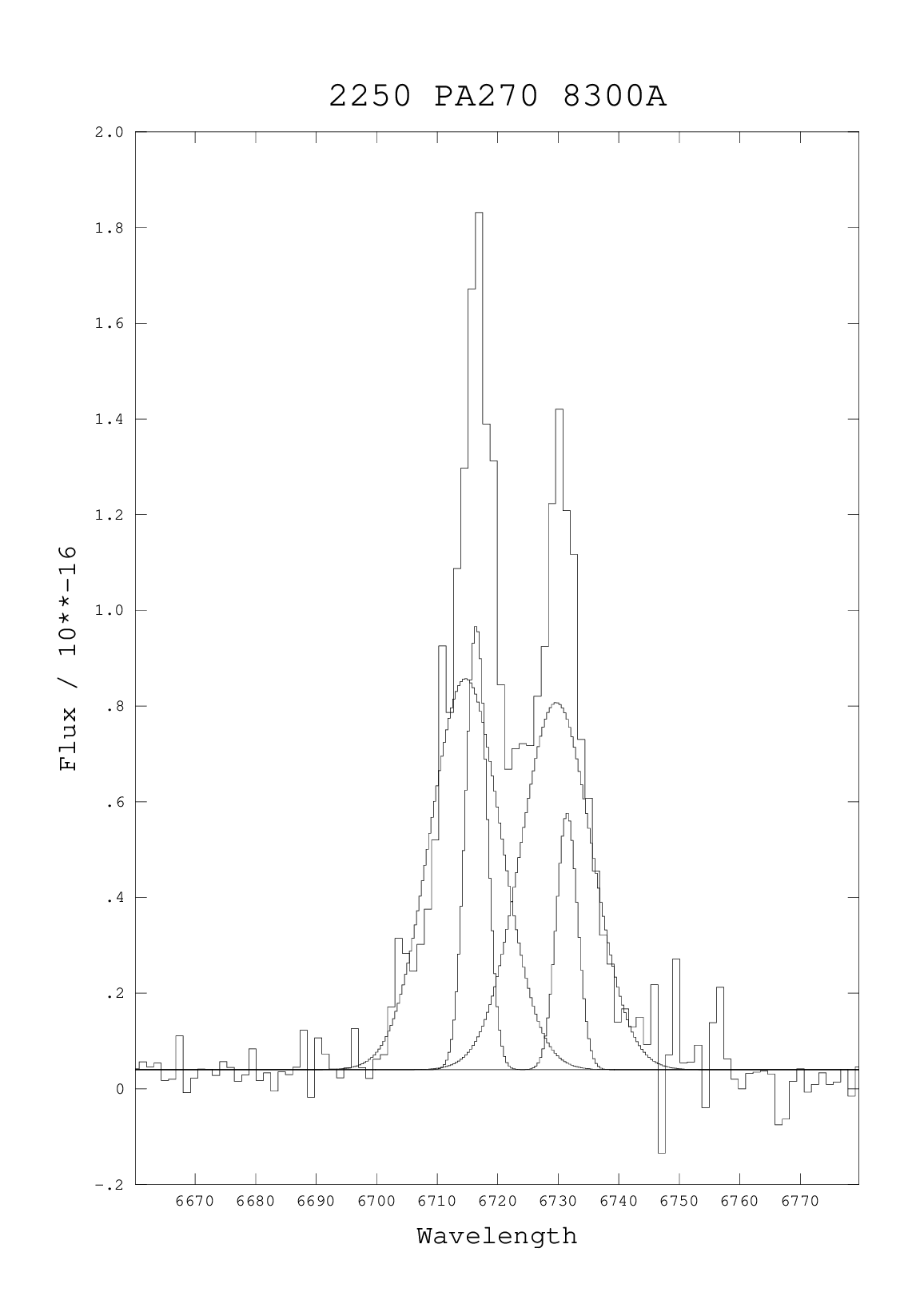}
\vspace{3in}
\caption{{\it The fit to the [SII]$\lambda\lambda$6716,6731 doublet}. 
Left: Solid smooth line represents the fit to the data.  Right: Solid smooth
lines represent the individual components (narrow and broad).}
\end{figure*}
  	
\subsubsection{The densities}

 	 We have attempted to use  the [SII]$\lambda\lambda$6716,6731  
doublets to calculate the density (Osterbrock 1989) of the broad and the narrow 
components. 
We show in Fig.~9  the results
of the   fit.

We obtained $\frac{[SII]\lambda 6716}{6731}$=1.47$\pm$0.191 for the narrow component 
(consistent with low density limit), and
$\frac{[SII]\lambda 6716}{6731}$=1.13 $\pm$0.08 for the broad
component. These results provide tentative evidence that
the broad component has a higher
density ($n_e = (5.7_{-1.9}^{+2.6})\times10^{8}$ m$^{-3}$, assuming $T_e$=30,000) than 
the narrow component, although
data with a higher signal-to-noise ratio  and a better sky subtraction 
will be required to confirm this result and provide a more accurate
estimate of the density contrast between the two components.

In order to provide a
further check on our results, we
have measured the total [SII]$\lambda\lambda$6716,6731 fluxes by
fitting only a single component to each line in the
blend. This leads to
$\frac{\lambda 6716}{\lambda 6731}$=1.24$\pm$0.06 and a density 
of $n_e= (1.9_{-0.6}^{+0.8})\times10^{8}$ m$^{-3}$ (assuming $T_e$=10,000)
 --- consistent with Clark {\it et al}. (1997),
who obtained $n_e = (1.7\pm0.5)\times10^{8}$ m$^{-3}$. Note that, if the narrow component 
has a low density --- as tentatively suggested
by our multi-Gaussian fits to the blends --- and taking into account that
the temperature of the broad component is higher, then $n_e\sim1.9\times10^8$ m$^{-3}$ 
represents a lower limit on the density of the broad component.

   \section{Discussion}

\subsection{The nature of the different kinematic components}

\begin{figure*}  
\includegraphics{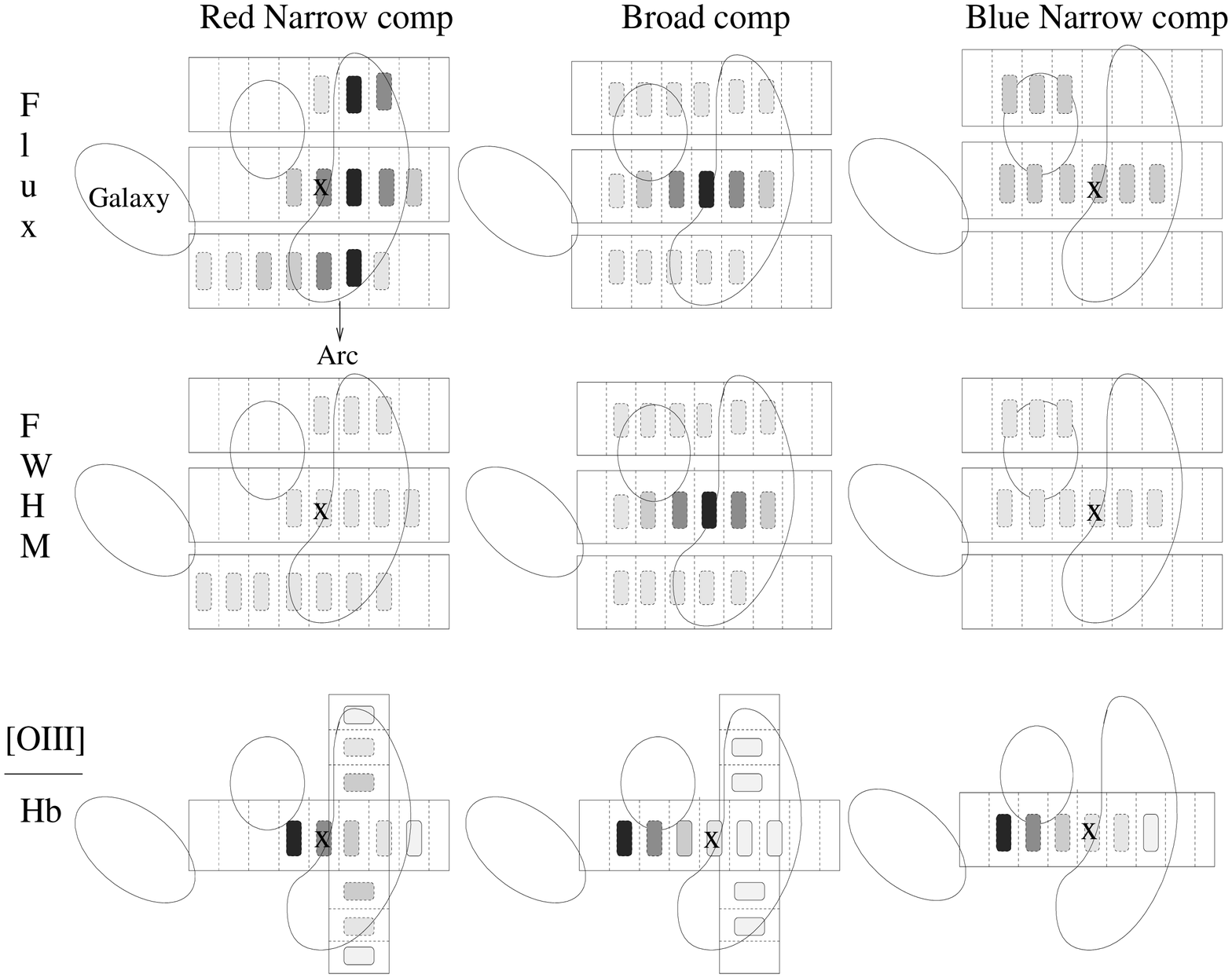}
 \vspace{6in}
\caption{Cartoon showing the spatial variation of the flux (top  panels), FWHM (middle 
horizontal panels) and ionization level (lower panels)
for the different kinematic components detected in PKS2250-41: Red narrow component 
(left panels), broad component  (middle vertical panels),
blue narrow component (right panels). The position of the hot spot has been indicated 
with an 'x'.}
\end{figure*}

	The cartoons presented in Fig.~10 show 2-D pictures of the Western arc, with the 
spatial variation of the flux, FWHM and
ionization level of the gas indicated. Darker squares indicate higher 
values. The relative intensity scale  is  approxi\-mately correct for each cartoon, 
but not from cartoon to cartoon. 
 ({\i.e.} the same colour intensity in a given cartoon implies same value, but not in 
two different cartoons). The cartoons summarize
the properties of the spatial variations of the different components in a more
visual way. Notice the very similar line flux curves, ionization
and kinematic properties of the red narrow component across the arc; 
the constancy of  
the flux and FWHM of the broad components
across the arc, except at the position of the hot spot; and the presence 
of the blue 
component  along the radio axis (and detached blob) with rather
constant flux and FWHM.  Notice also the increase in the ionization level for all the 
components as the distance to the nucleus of the host galaxy
decreases.

\vspace{0.15cm}

 \centerline{\it The broad component}
 
\vspace{0.15cm}

	 The similarities in the   properties of the broad component measured
in all spatial positions, except in the proxi\-mity of   the hot spot, suggest
that the broad components detected in all spectra  have a common
origin. This origin is clearly associated with   the interaction between 
the radio-emitting structures and the ambient gas because:
	
	- the large velocity width is incompatible with pure gravitational
motions at the position of the hot spot,  and is also rather large at the other 
positions;

	- the very low ionization level is as expected from the compression
effect of the shocks and/or the influence of shocks in the emission line
processes; 
	
	- the flux and line width   peak   at the position of the radio hot spot. Again, 
this is what we expect, since 
 the hot spot  is the working surface of the radio jet and the effects of the
interaction are expected to be stronger here.

 
	- the high electron
temperatures revealed by the [OIII] lines are consistent
with the idea that  the broad component is cooling gas
behind a shock front.  Although the density results are not conclusive, they suggest
that the shocked gas has been compressed.

	- the faintness of the HeII line also suggests the effects of shocks,
since this  property is predicted by
shock models
and similarly weak HeII lines are observed in other jet-cloud
interaction targets ({\it e.g.} 3C171, Clark {\it et al.} 1998)
 
	The possibility that the broad component is scattered light is
rejected since it is detected in the forbidden lines and the polarization level of the 
continuum is low 
(Dickson  {\it et al.} 1995).

The spatial extent of the broad component ($\sim$4.86 arcseconds or $\sim$30 kpc in the N-S 
direction) indicates that the
effects of the interactions reach regions as far as $\sim$15 kpc in the
direction perpendicular to the radio axis --- approximately the width of the
radio jet cocoon.   

\vspace{0.15cm}

   \centerline{\it The blue narrow component}
 
\vspace{0.15cm}

	The apparent association
of the blue narrow component with  the radio axis along PA0,
the abrupt disappearance of this component
at the edge of the radio lobe, and the drop of the flux at the
position of the hot spot suggest that this component may be 
affected by the  interaction. There is marginal evidence from the
[OIII](5007)/H$\beta$ ratio along PA0 (Fig.~3) that this component
has a lower ionization state than the red narrow component. If confirmed,
this result is consistent with the idea that the ionization of the blue
narrow component has been affected by the compression and/or ionization
effects of the jet-induced shocks.

Alternatively, this component may simply represent a separate velocity
system in the ambient gas of the host galaxy, with a higher density
or different distance from the nucleus resulting in a lower ionization
state. In this case, the apparent association with the radio axis along
the cross-cut slit position would
be coincidental. 
\vspace{0.15cm}

  \centerline{\it The red narrow component}

\vspace{0.15cm}

The emission  of the red narrow component 
is characteri\-zed by little variation in properties across
the arc, including the radio axis and  the position of
the radio hot spot.  The line widths are consistent with gravitational
motions  ($\sim$150 km s$^{-1}$). The ionization level is high relative
to the broad component. None of these properties present changes related
to the radio structures. Moreover, this
component is detected far beyond
the radio hot spot (and beyond the edge of the radio lobe).

	All of these properties  suggest that the red narrow component
 is emitted by ambient material which has not yet been shocked. 

	If the material emitting the red narrow
component had passed through the shocks and subsequently cooled
behind the shock fronts, we would expect a sudden decrease
in the ionization state of this component at the position of the edge of
the radio lobe, due to the compression effect of the shocks.
As the H$\beta$ line is too faint beyond the  radio lobe, we have
used the [OII]/[OIII] ratio to compare the ionization level beyond and
behind the edge of the radio lobe. We have extracted   a  spectrum
from a 1.6$\times$1.5 arcsecond aperture (2 pixels in the spatial direction) immediately 
behind
the edge of the radio lobe and another spectrum
from a  2.4$\times$1.5 arcsecond  aperture (3 pixels)
immediately beyond it. We obtain [OII]/[OIII]=2.1$\pm$0.2  behind  and 
2.0$\pm$0.3 beyond the edge of the radio lobe. Therefore, there is not
the sudden change in the ionization level that we would expect
if the red narrow component represents cooled post-shock gas.

	 Interestingly, 
the spatial distribution of this component
--- particularly the evidence for edge brightening along the slit position
perpendicular to the radio axis (Fig.~5) --- shows that it is associated
with the arc structure observed in the [OIII] emission line image
(see Fig. 1). The question then arises as to why this component has
an arc shape --- reminiscent of a bow shock --- yet there is no spectroscopic
evidence that the gas emitting the red narrow lines has been
disturbed by the shocks. We will return to this question in section 4.4.1. 

\vspace{0.2cm}

{\it Therefore, we have resolved kinematically the emission from the
shocked gas and the ambient gas in PKS2250-41.  Both gaseous components
present different kinematic and physical properties.}

\vspace{0.2cm}

We note striking similarities between PKS2250-41 and two other jet-cloud
interactions candidates which have recently been studied in depth:
3C171 (Clark {\it et al.} 1998) and PKS1932-46 (Villar-Mart\'\i n {\it et al.} 1998). 
In the case
of 3C171, broad and narrow components are observed across the extent of
the  interaction and, just as we
observe in PKS2250-41, the broad component has a low ionization state
while the narrow component has a high
ionization state and extends beyond the radio structure. The HeII$\lambda$4686 line 
(integrated flux) 
is also  weak. Although the case
for a strong jet-cloud interaction in PKS1932-46 is weaker, we also find
evidence in that object for a narrow, high ionization component which extends
well beyond the radio lobe, and a broader, lower ionization component which may
be related to a jet-cloud interaction. Remarkably, this similarity in the
kinematic/ionization properties persists despite a wide range in the 
emission line morphologies of the individual sources.

%
%

\subsection{The scenario}

	We now propose a scenario to explain the origin and pro\-perties of
the different kinematic components resolved in PKS2250-41. This scenario
could be applied to other radio galaxies in which jet-cloud interactions 
are taking place.

  	Hydrodynamical models of an advancing jet in a gaseous medium show
that  a bow shock is created that expands transversely to the jet axis 
({\it e.g.}  Hartigan {\it et al.} 1987, Taylor {\it et al.} 1992) (see Fig.~11
based on Bicknell {\it et al.} 1997).

	The models also predict that the shocked   gas is  
first  heated to temperatures beyond 10$^6$K and cools slowly in a   
phase ruled by the decreasing pressure (e.g. Ferruit {\it et al.} 1997). When the 
gas cools to a temperature of a few 10$^5$K  a catastrophic cooling occurs. The 
temperature drops very quickly, while the
density increases.  When the gas cools down to a few $\times$10$^4$K we will be able to 
observe the line emission. The post-shock gas 
contains a significant range of gas temperatures and the [OIII]$\lambda$4363 line 
luminosity is dominated by emission from the 20,000--30,000~K
zone.
This is consistent with the   high
observed temperatures in the broad component and the possible   compression suggested by 
the low ionization state and 
the density measurements.

\begin{figure}   
\includegraphics{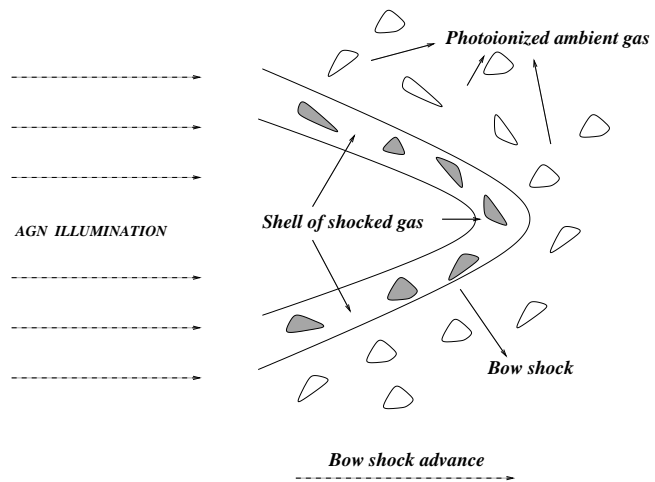}
\vspace{3.5in}
\caption{{\it The bow shock model}. (Bicknell {\it et al.})}
\end{figure} 

 	The gas passing through
the shock front will be acce\-lerated and its   kinematics
will be perturbed. The effects are likely to be stronger at the position of the hot 
spot,
the working surface of the radio jet. This is also what we observe (large FWHM with 
maximum value at the position of the hot spot).

Therefore, the properties of the broad component  are consistent with it
representing gas which has passed through
the shock front and been accelerated, heated and compressed.     Our results suggest 
that we are observing the gas   while the
dramatic cooling  is happening (it  is hotter than the ambient gas). 

The blue
narrow component is reminiscent of  linear structures associated
with the jets in some Seyfert galaxies. 
Such structures are characterized by high
ionization level compared to the gaseous cocoon around the radio lobes 
(Capetti {\it et al.}, 1996). The similarities with the properties
of the blue narrow component in PKS2250-41 suggest a similar origin, although
the nature of such linear features is not understood with any certainty.

	On the other hand, the properties of the red narrow component 
are consistent with the idea that it is emitted by warm
gas which has not yet been shocked. It reveals the properties of the
underlying quiescent gas. However, uncertainties remain concerning how this
component is
ionized (see \S4.4.1).

\subsection{The acceleration mechanism}

	We have measured line widths 
of up to $\sim$900 km s$^{-1}$ (FWHM) in PKS2250-41. 
The real velocities could be much larger, since, given the
relatively weak radio core (Clark et al. 1997), it is likely
that the jet is moving close
to the plane of the sky, and any systematic
motions will be induced in that plane. Line widths of up to
several thousands km s$^{-1}$ have also
been measured on the EELR of many high redshift radio galaxies ({\it e.g.} Mc.Carthy 
{\it et al.} 1996,
R\"ottgering {\it et al.} 1997).

\begin{figure*}
\includegraphics{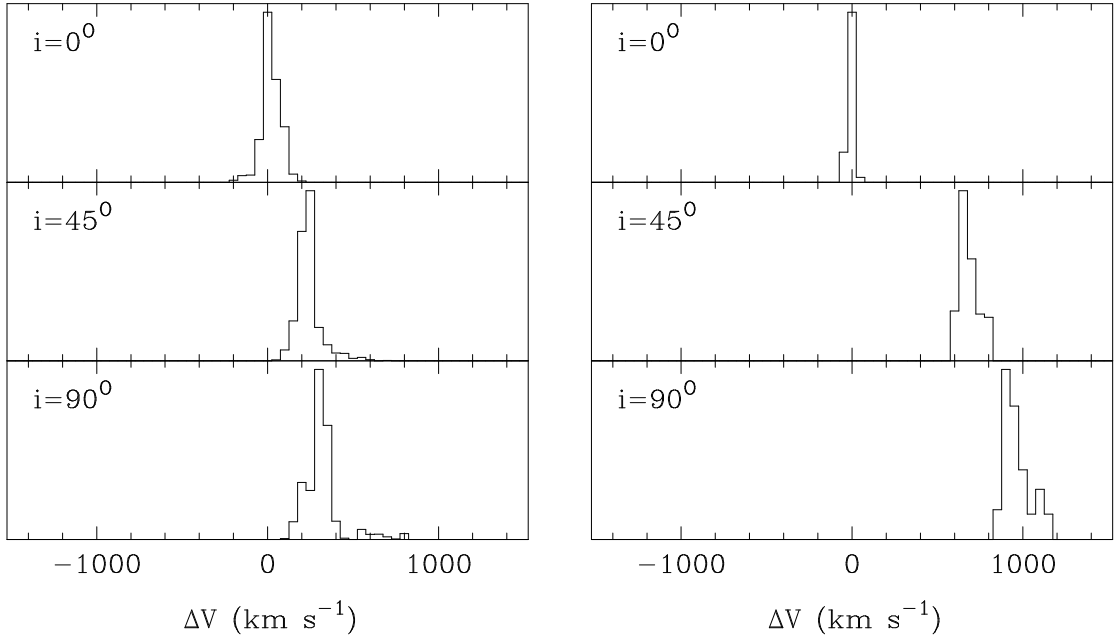}
\vspace{4in}
\caption{Velocity distributions of material along the line of sight for a
cloud (10pc diameter) that is impacted by an expanding bowshock associated with the 
radio
plasma. Left and right panels correspond to two stages during the disruption 
($t = 2 \times 10^4$ yr on the left and 
$ 4 \times 10^4$ yr on the right). The histograms show the total amount of mass, or 
relative column
densities, of all the material with temperature $T < 10^6$ K along the line of
sight in each velocity bin, with three
inclinations of the shock velocity vector to the line of
sight ($i$ represents the orientation of the entrained cloud material with
   respect to the plane of the sky). The parameters of the cloud, the ambient medium
and the bowshock are appropriate for a rich group of galaxies
(consistent with the X-ray properties of PKS2250-41: Clark et al.
1997): an
initially homogeneous, spherical cloud of temperature $T \sim 15,000K$, in
pressure balance with an ambient medium of $T \sim 10^7$K and pressure
$P \sim 4 \times 10^{-12}$ dyn cm$^{-2}$, is impacted by a shock moving at
5000 km s$^{-1}$. The simulation was carried out using the Zeus-3D code, in a
3-dimensional volume of 192$\times$64$\times$ 64 cells, with the cloud
diameter initially resolved by 24 cells.
}
\end{figure*}

One of the big challenges for models of  interactions between the radio and optical
structures is
to explain how the warm gas is accelerated to such large velocities.  The
jet propagates through a two phase medium, with warm (T$\sim$10$^{4}$ K) clouds
embedded in a hot (T$\sim$10$^{7}$ K) ambient phase, thus corresponding to a
density contrast of $\sim 1000$ if the clouds are in pressure balance with the
hot phase. There are several possibilities for accelerating the warm gas. These
include the following:
\begin{itemize}
\item {\bf Interaction between the cloud and the bowshock ahead of the jet
working surface}: in the absence of entrainment
processes (see below) the cores of the warm clouds will be accelerated to 
$V_w \sim V_s (\rho_h / \rho_w)^{0.5} $ (e.g. Klein {\it et al.} 1994) where, $V_w$
is the  final velocity of the warm clouds, $V_s$ is
the shock speed in the hot phase, $\rho_h$ is the preshock density of the hot phase,
and $\rho_w$ is the preshock density of the warm phase. Estimates of
hot spot advance speed (0.01$c$-0.1$c$, {\it e.g.} Scheuer 1995)  lead to
shock velocities in the range
3$\times$10$^3 < V_s <$3$\times$10$^4$ km $^{-1}$.
For warm clouds in pressure balance with a hot halo (T$\sim$10$^{7}$ K) the
density contrast is $\sim$1000 between the hot and the warm phases. Therefore,
the range of velocities of the accelerated warm clouds is 100-1000 km $^{-1}$.
However, noting that this is the maximum we would expect for the working surface of the
jet, that much of this velocity is likely to be directed in the plane
of the sky (the direction of the jet), and that the cocoon away from the
hotspot is likely to expand much more slowly than at the working surface, it
is unlikely that the bowshock acceleration is sufficient to explain the large
line widths observed in PKS2250-41 and other jet-cloud interaction candidates. 
\item {\bf Entrainment in the hot post-shock wind.}
An interesting possibility is that the clouds are entrained in the
hot wind behind the shock. Instabilities will arise at the interface between
the hot, fast flowing gas and the warm, much slower moving clouds. These 
instabilities will lead to entrainment and acceleration of warm material in
a hot wind (e.g. Stone \& Norman 1992;
Klein {\it et al.} 1994;
Dai \& Woodward 1994). 
Koekemoer (1998)
has recently completed a
grid of models in the specific context of clouds with high density contrast
($\sim 1000$) in the hot gas around radio galaxies, which are impacted
by shocks with properties typical of expanding radio cocoons. In Fig.~12 we
show results from one such simulation carried out in the specific context of
PKS~2250-41: a cloud initially at $T \sim 15,000K$ embedded in a hot interstellar medium 
with conditions appropriate to a rich group of galaxies is
impacted by a shock moving at 5000 km s$^{-1}$, expected to be typical of the
bowshock expansion velocity. The velocity profiles of the stripped gas (with
$T \leq 10^6$ K) are shown. Although this is a relatively simple scenario,
not yet taking into account time averaged profiles or different distributions
of cloud sized along the line of sight, it is nevertheless clear than the
velocities of order $\sim 1000$ km s$^{-1}$ can be produced
in this way. By integrating across the surface of the shock, for a range of
cloud properties and stages of interaction, it would clearly be feasible to
produce broad lines similar to those observed in PKS2250-41.

\item {\bf Entrainment of clouds in the turbulent boun\-dary layers between the
radio jet and the ISM or the
jet cocoon and the ISM} (e.g. Sutherland {\it et al.} 1993). This mechanism
could also produce large line widths and accelerate the clouds to
a significant fraction of the jet speed.
\end{itemize}

Therefore it is likely that, in addition to
the initial shock acceleration, an entrainment process is required to explain
the large widths of the broad emission lines. Higher
spatial resolution optical and radio observations will be required to determine
whether the gas is being entrained in the hot post-shock wind
or in the turbulent boundary layers of the radio jets.

 \begin{figure*}
\includegraphics{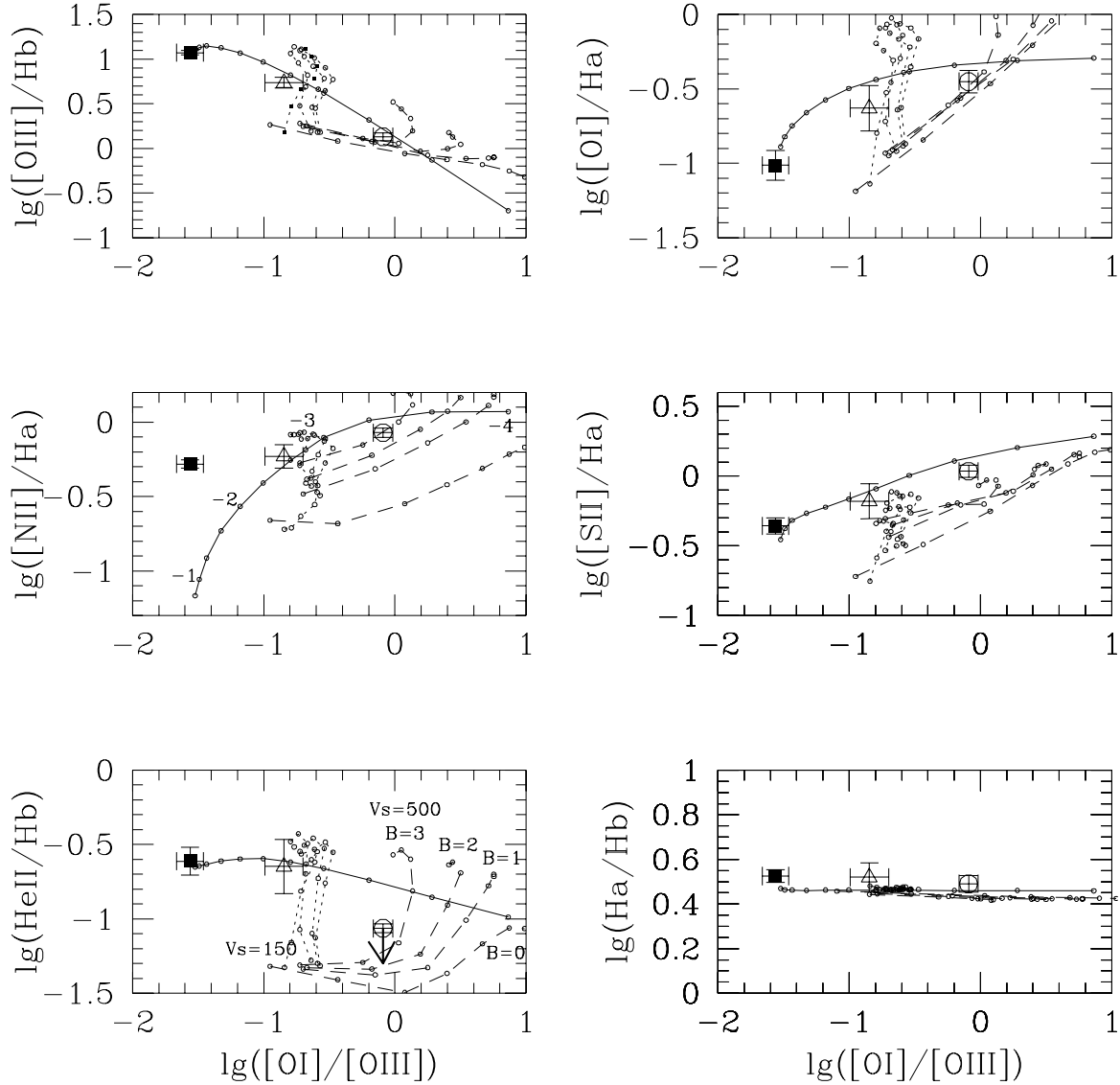}
\vspace{6in}
\caption{Diagnostic diagrams:  
 AGN photoionization models  --$>$ Solid line ($logU$ values are indicated in
the log[NII]/H$\alpha$ {\it vs.} lg([OI]/[OIII]) diagram). Shock models
(cooling gas)  --$>$ Long dashed lines. Shock+precursor models  --$>$ dotted
lines (values of the shock velocity $v_s$ and the magnetic parameter $B$ are 
indicated in the log(HeII/H$\beta$ {\it vs.} lg([OI]/[OIII]) diagram).
\hspace{3cm}Nucleus --$>$ Solid square. Broad component --$>$ open circle. Narrow 
component --$>$ open  triangle.}
\end{figure*}

\subsection{Ionization mechanism: shocks {\it vs.} AGN photoionization}

	Shocks can influence the ionization of the gas in two ways (Sutherland {\it et 
al.} 1993,
Dopita \& Sutherland 1996):
1) the initial heating of the gas as it passes
through the shock front, leading to collisional ionization
of the gas and, eventually, to forbidden line emission
as the gas cools radiatively; 2) the generation of a strong UV continuum 
in the hot post-shock gas
which can ionize both the precursor gas and also the gas
which has cooled radiatively behind the shock front. However, the
galaxy also 
contains an active nucleus which is likely to generate a hard continuum
capable of ionizing the gas.   Which mechanism  dominates the ionization
of the extended emission line gas in PKS2250-41: shocks or AGN photoionization?

Taken at face value, the high electron temperature measured for the 
broad component in the western arc provides strong evidence that 
the gas emitting the broad component is cooling behind
a shock front. Such high temperatures are predicted by models for shocked,
cooling gas, but are difficult to explain in terms of conventional, radiation
bounded AGN photoionization models. Although photoionization 
models which include 
matter-bounded clouds, or clouds photoioni\-zed by the hard continuum which
has been partially absorbed by a matter-bounded component, can produce high
electron temperatures (Binette {\it et al.} 1997), such models cannot explain a 
temperature
as high as $T_e > 20,000$K and a low HeII(4686)/H$\beta$ ratio simultaneously.

In order to check the consistency of our
results with the ionization models in greater depth, 
we have compared the measured values for the line ratios of
the different gaseous components in PKS2250-41 with the prediction of both AGN 
photoionization
and shock models. The results are shown in the diagnostic diagrams 
presented in Fig.~13.      
We have used the multi-purpose photoionization code MAPPINGS~I (Binette {\it et~al.} 
1993a,b)
to generate the AGN model predictions. We have assumed a density of 100 cm$^{-3}$
at the illuminated face of the clouds, solar abundances, and a power-law spectral
index for the photoionizing continuum of $\alpha = 1.5$ ($F_{\nu}\propto 
\nu^{-\alpha}$).
The results are presented as continuous lines in Fig.~13
as  sequences in the ionization parameter $U$, defined by

	$$ U = \frac{1}{cn_H} \int_{\nu_0}^{\infty} {\frac{\phi_{\nu}}{h\nu} d\nu}$$

where $c$ is the speed of light, $n_H$ is the density of the gas in the front layer, 
$\nu_0$ is the Lyman limit frequency, and 
$\phi_{\nu}$ is the monochromatic ionizing energy flux  impinging
on the slab.    
Note that this same sequence is able to reproduce the general trend defined
by the optical line ratios of low redshift radio galaxies (Robinson {\it et al.} 1987).

	For the shock models we have used
the published results of  Dopita and Sutherland (1995), whose models
include both the gas cooling behind the shock front, and the gas
photoionized by the hot post-shock gas (the photoionized
precursor).    
The two main parameters which
influence  the predicted spectrum  are the velocity of the shock ($v_s\subset$[150,500] 
km s$^{-1}$) and the magnetic
parameter ($B\subset$[0,4]  $\mu$G cm$^{-3/2}$). Adopted density
 is n(H)=1cm$^{-3}$ and solar abundances.	
The spectrum predicted for the cooling gas is   presented in the diagnostic diagrams in 
Fig.~13 as long dashed lines. Each sequence corresponds to a 
fixed $B$ value and changing $v_s$. 
 The spectrum predicted for the precursor is presented as dotted lines  
($v_s\subset$[150,500] km s$^{-1}$
and fixed $B=$1  $\mu$G cm$^{-3/2}$).

	Fig.~13 shows that  the broad and narrow component occupy very
different positions in the diagnostic diagrams, due mainly to a lower
ionization level of the broad component. Although the line ratios for the broad 
component are reproduced by
the  AGN photoionization models on many of the diagrams, an 
important problem of AGN models is the failu\-re to explain
the high electron temperatures 
and the faintness of the HeII$\lambda$(4686) line for the broad component.
We plot in Fig.~14 a new diagnostic diagram: [OIII]$\lambda$5007/4363 {\it vs.}  
HeII/H$\beta$. This diagram  clearly distinguishes 
the narrow and the broad  
components.  The AGN photoionization models do not appear in 
this diagram at all because they predict high
[OIII]5007/4363 ratios ({\it i.e.} corresponding to low electron
temperatures), which fall above the top of the diagram. Taking
Figures 13 and 14 together it is clear that the line ratios of the 
broad component are entirely consistent with the idea that  
this component represents collisionally ioni\-zed gas which is
cooling behind a shock front; the shock velocities able to reproduce the
position of the broad component in the diagnostic diagrams are in the
range 200-300 km s$^{-1}$. The AGN photoionization models fail
to produce such a good overall fit to the emission line spectrum.   
Thus we have strong evidence that the jet-induced shocks
not only compress, but 
also ionize, the warm clouds ({\it i.e.} the jets have a significant
energetic input).

\begin{figure}  
\includegraphics{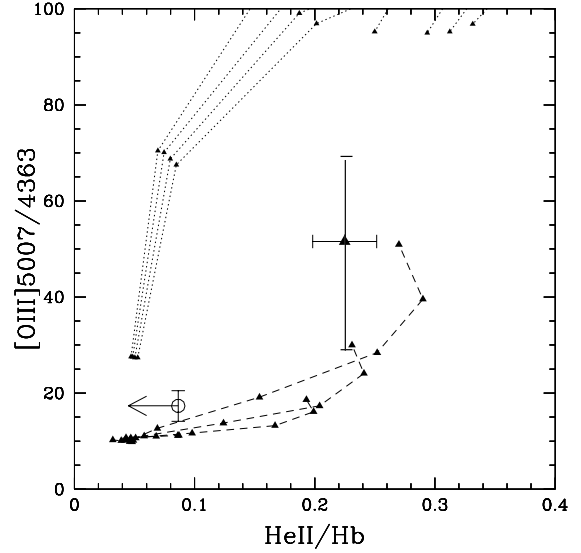}
\vspace{3.5in}
\caption{{\it The temperature}: [OIII]$\lambda$5007/[OIII]$\lambda$4363
{\it vs.}  HeII/H$\beta$.  The broad   (open circle) and the narrow
(solid triangle) components are clearly separated in this diagram. 
Dotted lines: shock+precursor models. Dashed lines: shock models.
AGN photoionization models lie outside the diagram. They produce too 
high [OIII]5007/4363.}
\end{figure} 

In contrast to the situation for
the broad components, the AGN and shock precursor 
photoionization models produce  
good fits to
the line ratios of the narrow components on most of the diagrams. Any
remaining discrepancies between the measured and predicted 
[OIII]5007/4363 ratios for the narrow components  can be resolved by adding a
matter-bounded component (Binette {\it et al.} 1997). 

\subsubsection{The ionization of the red narrow component}

We noted in section 4.1 that there is an apparent inconsistency
between, on the one hand the association of the red narrow component
with the arc structure, and on the other the lack of
clear spectroscopic evidence that the material emitting this
component has been shocked. One possible explanation --- consistent
with the emission line spectrum of this component --- is that
the red narrow component represents shock-photoionized precursor
gas.

The shock photoionization process will be most efficient
when the precursor gas is  situated in ~the clouds that are being
shocked, since the precursor gas will then absorb a large fraction of
the ionizing photons generated in the shocks. As the fast shocks being 
driven through the hot ISM by the radio source engulf denser clumps of
warm gas, slower shocks will be driven through
the warm clouds (see section 4.3). The X-ray photons generated in the hot gas 
immediately behind the shocks in the warm clouds will not be efficient at photionizing
the warm gas, however, provided that the gas cooling time ($t_{cool}$) is short compared 
with the shock crossing time ($t_{cross}$), a cooling zone will form downstream of the 
shock which will
emit lower energy photons, capable of ionizing the precursor
gas efficiently. Thus we require
$t_{cross} > t_{cool}$ for efficient precursor ionization. Using the 
appro\-ximate formula for the cooling time given in section 2.3 of Klein {\it et al.}
(1994) this leads to the following cons\-traint on the diameter ($D$) of 
warm clouds:
\begin{displaymath}
D > 1.8 \left(\frac{n_0}{10^6} \right)^{-1} 
\left(\frac{V_w}{200 \> \rm km \> s^{-1}}\right)^4
\hspace{0.5cm} pc
\end{displaymath}
where $n_0$ is the precursor density (in units of $m^{-3}$), and $V_w$ is the
velocity of the slow shock through the warm cloud. Thus, for a precursor
density typical of warm clouds in pressure equilibrium with the hot
X-ray halo of a group of galaxies ($n_0 \sim 1\times10^6 m^{-3}$) and a shock speed
consistent with the measured emission line ratios of the broad component
($V_w \sim 200$ $km \> s^{-1}$), we require $D > 1.8$ pc for efficient precursor 
photoionization. Although we have no direct indication of the diameters of
the warm clouds interacting with the radio components in PKS2250-41, it is notable that 
typical spiral galaxies contain populations of molecular
cloud complexes with diameters that are orders of magnitude larger
than this limit ({\it e.g.} Mihalas \& Binney 1981, Chapter 9). Therefore, it is likely 
that the PKS2250-41 
system will contain at least some clouds for which the precursor photoionization
process envisaged above is feasible.   

We now 
estimate the mass flow rate  required
to maintain the narrow $H\beta$ luminosity ($L_{H\beta}$). 
For a strong shock, the H$\beta$ luminosity
of the photoionized precursor gas is related to the shock velocity through
the warm clouds
($V_w$) and the mass flow rate through the shock  ($\dot{M}$) by the following
equation (adapted from equation 4.4 of Dopita \& Sutherland 1996):

\begin{displaymath}
L_{H\beta} = 1.45\times10^{31}\left(\frac{V_w}{200 \> \rm km \> s^{-1} }\right)^{1.3} 
\left(\frac{\dot{M}}{M_{\odot} yr^{-1}}\right)
\hspace{0.5cm} W
\end{displaymath}

For our chosen
cosmology, we estimate a total H$\beta$ luminosity for the red narrow
component emitted by the arc of $L_{H\beta}$ = 2.2$\times$10$^{34}$ $W$.
Substituting this into the expression for the $H\beta$ luminosity
we obtain:

\begin{displaymath}
\dot{M} = 1.5\times10^{3} \left(\frac{V_w}{200 \> \rm km \> s^{-1}}\right)^{-1.3}
\hspace{0.5cm} M_{\odot} yr^{-1}.
\end{displaymath}

To put this mass flow rate into context, if the
radio source requires $\sim$10$^6$ yr to traverse the region of the
arc, and $V_s \sim200$ $km s^{-1}$, the total amount of material 
``consumed'' in the
shock will be $1.5\times10^{9}$ M$_{\odot}$ --- comparable with the total
HI content of a typical spiral galaxy. Although large, this is consistent
with the idea that the phenomena we observe in PKS2250-41 are the result
of a direct interaction between the radio-emitting components and a companion galaxy. 
Note also that there is more than enough mechanical
energy in the jets to power the emission line regions (see Clark {\it et al.} 1997).

As well as the red narrow component directly asso\-ciated with the arc,  we
also see a more extended red narrow component which extends well beyond
the western radio lobe. Given its spatial extent, this component is unlikely
to be emitted by clouds which have been engulfed by the fast shock
driven through the hot ISM by the radio components. Nonetheless, it remains
possible that this component is photoionized by photons from
the warm, shocked clouds
behind the shock front, provided that the covering factor of the extended
component is large enough to intercept a significant fraction of ionizing
photons generated by the shocks. 
However, a potential problem with this scenario
is that the
effective ionization parameter of the shock continuum will be much larger
for the warm clouds in the arc than for the more extended warm clouds, yet
the measured ionization state for these two components is similar (see
section 4.1). While it is possible to get around this problem by assuming
that the more extended narrow component is associated
a lower density ISM than the narrow
component emitted by the arc, it seems unlikely that the density
gradient in the ambient ISM 
would exactly compensate for the change in the flux density
of the shock continuum with distance from the shock front.

In view of this potential problem, we should also
consider the alternative possibility: that the emission line arc represents
the {\it intrinsic} structure of the AGN-photoionized
ambient gas {\it i.e.} the observed arc-like morphology is not a
consequence of the shock, and the resemblance to a bow shock
is misleading. Arc-like structures in the ISM are readily produced as a 
consequence of galaxy mergers and interactions, and PKS2250-41 shows
morphological evidence for interactions with at least one companion
galaxy (Clark et al. 1997). However, it would seem a remarkable coincidence
if, purely by chance, such a merger remnant exactly circumscribed a 
radio lobe. Furthermore, arc-like structures circumscribing the radio
lobes have been observed in other radio galaxies with
jet-cloud interactions ({\it e.g.} PKS1932-464, Villar-Mart\'\i n {\it et al.} 1998 and 
Coma A, Tadhunter {\it et al.} in preparation).

Thus, no explanation for the ionization and distribution of the
red narrow component seems entirely satisfactory, and this issue clearly
warrants further investigation.

\subsection{Shocked and ambient gas in high redshift radio galaxies}

 	PKS2250-41   shares many similarities with high redshift radio galaxies.
Therefore, what we learn studying this object will help us to understand
the observed properties of very distant radio galaxies.
Like many high-z radio galaxies, PKS2250-41 shows  perturbed kinematics and a close 
correlation between the EELR and the radio 
structures.  McCarthy \& van Breugel (1989)  noted that the brightest EELR  in
radio galaxies at z$\geq$0.7 tend to be associated with
the radio lobe closest to the nucleus of the host galaxy 
which moreover, is systematically more depolarized   (Pedelty {\it et al.}  1989, Liu \& 
Pooley
1991). PKS2250-41  also shows these asymmetries (Clark {\it et al.} 1997) 
which may arise, as proposed for high-z radio galaxies, as
a consequence of an asymmetric gaseous environment surrounding the host galaxy. 

We have shown that the strong interactions between the radio jet and the ambient gas  
imprint special
features in the observed properties of the ionized gas, which differ from
the ambient (non shocked) gas properties.	{\it Do we see these 
features in  HzRG?}.
 
	At least some HzRG show, in addition to the high
velocity gas associated with the radio structures,  large halos
of ionized gas   extending far beyond the radio structures. Interes\-tingly,
the kinematics of this gas are rather quiescent compared to the inner 
structures. Examples include   1243+036 ($z=$3.6, van Ojik {\it et al.} 1996),
3C368 ($z=$1.1, Stockton {\it et al.} 1996) and 4C41.17 ($z=$3.8, Dey {\it et al.} 1997, 
Chambers {\it et al.} 1990).  The high velocity gas in both 3C368 and 4C41.17 presents
low ionization level compared   to the low velocity gas.  

 	1243+036, 3C368, 4C41.17 are   examples of   distant radio galaxies
which show the presence of two types of ionized gas.  The similarities
with PKS2250-41 suggest that the na\-rrow line gas (at least the gas detected beyond the 
radio structures)  is probably quiescent  ambient gas  which
has not interacted with the radio structures  and is ionized by  the AGN continuum
and/or the shock UV continuum. On the other hand, the high
velocity gas is  shocked gas which has been heated, compressed and kinematically 
disturbed by the shocks.  

	 Therefore it seems likely that
the ambient and shocked gas have also been resolved in some of
the high-z objects, and
they present differences in their kinematics and ionization properties 
consistent with the results obtained for PKS2250-41. A similar  kinematic study  as 
described in this paper will be very fruitful in these objects.
It will allow us to  isolate  the  non-shocked ambient gas and study the 
properties of the gaseous halos in elliptical galaxies in the young Universe
as well as to understand the way the observed properties are determined by
the interaction with the radio-emitting plasma.

\section{Summary and conclusions}

	We have studied the effects of the interaction between the ambient gas
and the radio-emitting plasma in PKS2250-41. The results of this research provide 
clues for understanding the processes which determine the observed
properties of the majority of very distant radio galaxies and to determine
the r\^ole played by the interactions between the radio and optical structures in active 
galaxies in general.

	We have resolved kinematically the emission from the gas 
which has passed through the shocks induced by the radio plasma and the emission from 
the ambient precursor gas. The properties of the two main gaseous components show
marked differences.

Our analysis shows that the effects of the interac\-tions can spread
at least 15 kpc perpendicular to the radio axis --- across the full
extent of the radio lobes.
The interac\-ting gas presents broad line widths (FWHM$\geq$900 km s$^{-1}$),  a low 
ionization level and weak HeII$\lambda$4686
compared to the ambient gas.      
We have presented the first evidence that the  gas emitting the 
broad component
is   hotter ($T\sim$30,000 K) than the ambient gas ($T\sim$15,000 K).  
All of these properties, together with the morphological association
with the radio structure, are consistent with
the broad component being emitted by gas which has been heated
and collisionally ioni\-zed by a jet-induced shock and is now radiatively
cooling behind the shock front. This represents compelling
evidence that jet-cloud interactions
not only compress and accelerate the interacting
warm clouds, but also ionize them.

	The ambient gas is characterized by a 
relatively high ionization level and
a more quiescent kinematics, consistent with gravitational motions. The properties of 
this ambient gas can be explained in terms of
pure gravitational motions and photoionization by a 
hard continuum source. It is not
yet clear whether this component is photoionized by the AGN or
by UV photons emitted
by the gas cooling behind the shock front. The arc shape
circumscribing the 
radio lobe suggests that the second mechanism is dominant, although a large
mass flow rate through the shock is required.
 
	The similarities between the kinematic properties of PKS2250-41 and
some  high-z radio galaxies suggest that the
ambient and the shocked gas have also been
resolved in the more distant sources, while the similarities between
PKS2250-41 and other jet-cloud interaction radio galaxies (3C171, PKS1932-464)
in spite of their very different morphology, suggest
that  jet-cloud interactions  have rather constant effects on the physical
and kinematic properties of the emission-line gas in powerful radio galaxies.

\section*{Acknowledgments}

The authors thank the referee, Dr. G. Bicknell, for a careful study of the paper
and for providing useful comments which
helped to improve it.
We would like to thank Bob Fosbury for reading the manuscript and contributing
useful comments. Thanks to L. Binette for his code MAPPINGS I, that we used to
build the photoionization models presented here.
M.Villar-Mart\'\i n acknowledges 
support from PPARC.

\end{document}